\documentclass[jgr]{agujournal2019}
\usepackage{url} 
\usepackage{lineno}
\usepackage[inline]{trackchanges} 
\usepackage{soul}

\usepackage{tikz}
\usepackage{algorithm}

\newcommand{\der}[2]{\frac{\partial #1}{\partial #2}}
\newcommand{\dif}[2]{\partial_{#1} #2}

\newcommand{\rbar}[2]{\left. #1 \right|_{#2}}

\newcommand{\olv}{\overline{v}}
\newcommand{\ols}{\overline{\sigma}}
\newcommand{\olJ}{\overline{J}}
\newcommand{\olM}{\overline{M}}
\newcommand{\ola}{\overline{a}}
\newcommand{\damatder}[1]{\frac{\Bar{d} \; #1}{\Bar{dt}}}
\newcommand{\sump}{\sum_{p=1}^{n_p}}
\newcommand{\sumi}{\sum_{i=1}^{n_g}}
\newcommand{\sumj}{\sum_{j=1}^{n_g}}

\graphicspath{{./Figures/}}
\usepackage[toc,page]{appendix}
\usepackage[T1]{fontenc}
\usepackage{url}
\usepackage[utf8]{inputenc}
\usepackage{lmodern}
\usepackage[english]{babel}
\usepackage{graphicx}
\usepackage{amssymb}
\usepackage{graphicx}
\usepackage{float}
\usepackage{caption}
\usepackage{subcaption}
\usepackage{amsfonts}
\usepackage{caption}
\usepackage{array,multirow,makecell}
\usepackage{amsthm}
\usepackage{natbib}
\usepackage{bm}
\usepackage{adjustbox}
\usepackage{romanbar}
\usepackage[intlimits]{amsmath}
\usepackage{amsmath}
\usepackage{srcltx}
\usepackage{array}
\usepackage{soul}
\usepackage{color}
\usepackage{algpseudocode}

\definecolor{darkWhite}{rgb}{0.94,0.94,0.94}

\draftfalse

\journalname{Journal of Advances in Modeling Earth Systems (JAMES)}

\begin{document}

%
%


\title{A Depth-Averaged Material Point Method for Shallow Landslides: Applications to Snow Slab Avalanche Release}

%
%




\authors{Louis Guillet\affil{1}, Lars Blatny\affil{1}, Bertil Trottet\affil{1}, Denis Steffen\affil{1}\\ and Johan Gaume$^{2, 3, 4*}$}

\affiliation{1}{School of Architecture, Civil and Environmental Engineering, EPFL, Lausanne, Switzerland}

\affiliation{2}{Institute for Geotechnical Engineering, ETH Zürich, Zürich, Switzerland}

\affiliation{3}{WSL Institute for Snow and Avalanche Research SLF, Davos, Switzerland}

\affiliation{4}{Climate Change, Extremes, and Natural Hazards in Alpine Regions Research Center CERC, \\ \vspace{-0.1cm}Davos Dorf, Switzerland}





\correspondingauthor{Johan Gaume}{jgaume@ethz.ch}




\begin{keypoints}
\item Development of a depth-averaged Material Point Method (DAMPM) for shallow landslides
\item Verification of the method based on analytical formulation
\item Application to snow slab avalanche release
\end{keypoints}

%
%

%
%


\begin{abstract}

Shallow landslides pose a significant threat to people and infrastructure. While often modeled based on limit equilibrium analysis, finite or discrete elements, continuum particle-based approaches like the Material Point Method (MPM) have more recently been successful in modeling their full 3D elasto-plastic behavior. In this paper, we develop a depth-averaged Material Point Method (DAMPM) to efficiently simulate shallow landslides over complex topography based on both material properties and terrain characteristics. DAMPM is an adaptation of MPM with classical shallow water assumptions, thus enabling large-deformation elasto-plastic modeling of landslides in a computationally efficient manner. The model is here demonstrated on the release of snow slab avalanches, a specific type of shallow landslides which release due to crack propagation within a weak layer buried below a cohesive slab. Here, the weak layer is considered as an external shear force acting at the base of an elastic-brittle slab. We validate our model against previous analytical calculations and numerical simulations of the classical snow fracture experiment known as Propagation Saw Test (PST). Furthermore, large scale simulations are conducted to evaluate the shape and size of avalanche release zones over different topographies. Given the low computational cost compared to 3D MPM, we expect our work to have important operational applications in hazard assessment, in particular for the evaluation of release areas, a crucial input for geophysical mass flow models. Our approach can be easily adapted to simulate both the initiation and dynamics of various shallow landslides, debris and lava flows, glacier creep and calving.

\end{abstract}

\section*{Plain Language Summary}

Shallow landslides represent a significant threat to people and infrastructure. In mountainous regions, snow slab avalanches are a particular type of shallow landslide responsible for numerous casualties and important damage. Such an avalanche releases due to crack propagation within a weak layer buried below a cohesive slab. Here, we develop a depth-averaged Material Point Method (DAMPM) to simulate efficiently the release of slab avalanches over complex topography based on snow properties and terrain characteristics. DAMPM is an adaptation of the Material Point Method with classical shallow water assumptions. The weak layer is considered as an external shear force acting at the base of the slab. The model is validated based on theoretical and numerical analysis of a state-of-the-art snow fracture experiment. Then, large scale simulations are conducted to evaluate the shape and size of avalanche release zones over different topographies. Given its low computational cost, we expect our model to have operational applications in hazard assessment, especially for the evaluation of the avalanche release size which is an important quantity for avalanche forecasting and management. The model can be easily adapted to simulate the initiation and dynamics of other processes such as debris and lava flows, glacier creep and calving.

%
%

%


%
%
%
%

\section{Introduction}


A snow avalanche consists of fast gravitational flow of a snow mass which poses a threat to snow enthusiasts and infrastructure in mountainous regions \citep{Ancey2006,Mcclung2006}. Among different avalanche types, dry-snow slab avalanches are responsible for most avalanche accidents and damage \citep{Schweizer2003}. Such an avalanche releases due to the failure of a weak layer buried below a cohesive snow slab \citep{Schweizer2016}. Although a lot of progress has been done regarding the understanding of slab avalanche release processes, including the onset and dynamics of crack propagation \citep{gaume2013, Bobillier2021, Trottet2022}, the evaluation of the size of avalanche release zones still remains a major issue. This difficulty impedes avalanche forecasting and hazard mapping procedures which rely on the avalanche release size as input.



Previous work tackled this obstacle through various modeling approaches. In particular, \cite{Fyffe2004}, \cite{Failletaz2006} and \cite{Fyffe2007} developed cellular-automata to compute the statistical distribution of the release zone area. These models succeeded in reproducing power-law distributions reported based on avalanche measurements \citep{Mclung2003, Fyffe2004}.  \cite{veitinger2016} proposed a fuzzy logic model that allows the identification of avalanche release areas on complex terrain based on topographical indicators (such as slope angle and curvature), forest indicators as well as roughness evolution induced by snow cover (progressive smoothing effect). Dynamic crack propagation in snow was simulated in 2D (depth-resolved) based on the Discrete Element Method \citep{Gaume2015} and the Finite Element Method \citep{Gaume2015a}. \cite{GaumeReuter2017} and \cite{Reuter2018} later proposed novel methods to describe snow instability by combining limit equilibrium and finite element simulations.  More recently \cite{Zhang2022} proposed a depth-averaged finite difference method for the initiation and propagation of submarine landslides. Yet, so far, approaches combining a mechanical release model, variable material properties and complex topography are still scarce.



In these lines, advanced numerical models were recently developed. \cite{Gaume2018} proposed a Material Point Method (MPM) and constitutive snow models based on Critical State Soil Mechanics to simulate in a unified manner, failure initiation, crack propagation, slab fracture, avalanche release and flow, at the slope scale. 
MPM is a Eulerian-Lagrangian particle-based method initially developed by \cite{SULSKY1994}. 
Due to its ability to handle processes including large deformations, fractures and collisions, this elegant hybrid method found great interest over the last two decades, both in geomechanics, e.g., for the modeling of fluid-structure interaction~\citep{york1999}, porous media micromechanics~\citep{blatny2021, blatny2022}, granular flows~\citep{Dunatunga_2015}, snow avalanche release~\citep{Gaume2019, Trottet2022}, snow avalanche dynamics~\citep{Li2021}, glacier calving~\citep{wolper2021}, debris flows~\citep{vicari2022}, landslides~\citep{soga2016} and rockslides~\citep{cicoira2022}, as well as in computer graphics~\citep{stomakhin, JIANG, SCHRECK2020, Daviet2016}. 
After its first application to snow slab avalanches, \cite{Gaume2019} analysed crack propagation and slab fracture patterns and reported crack speeds above 100 m/s on steep terrain. 
While the latter result was initially surprising, this motivated further analysis which later highlighted a transition in crack propagation regimes during the release process \citep{Trottet2022}. 
In fact, for short propagation distances, the failure in the weak layer occurs as a mixed-mode anticrack which propagates below the Rayleigh speed. 
In this case, simulated propagation speeds are similar to those reported in classical snow fracture experiments (Propagation Saw Test \citep{Alec2016}). 
For propagation distances, larger than the so-called supercritical crack length, a pure shear mode of crack propagation is reported with speeds higher than the shear wave speed, a process called supershear fracture which was previously reported in earthquake science. 
This transition, which occurs after a few meters suggests that a pure shear model should be sufficient to estimate the release sizes of large avalanche release zones.

Motivated by this new understanding and by the high computational cost of three-dimensional simulations, we developed a depth-averaged MPM for the simulation of snow slab avalanches release with a pure shear failure model for the weak layer. 
This model, inspired by the one developed in \cite{Abe16} for debris flows, extends the Savage-Hutter model \citep{Savage_Hutter} in the case of elastoplasticity. 
After presenting governing equation and depth-integration (Section 2), we validate the model based on simulations of the so-called Propagation Saw Test (PST) and compare numerical results to analytical solutions and 3D simulations (Section 3). 
Furthermore, we perform large scale simulations over generic and complex topographies and analyse the shape and size of avalanche release zones.

\section{Methods}

In this section, we first describe the governing conservation equations and derive their depth-integrated counterparts under the shallow water assumptions.
Then, in the context of snow slab avalanches, the constitutive elasto-plastic model for the slab as well as the weak layer-slab interaction are outlined.
Finally, a discretization of the equations are presented and the full depth-averaged Material Point Method (DAMPM) algorithm is summarised.

\subsection{Governing equations}
We denote $\bm{\sigma}$ the Cauchy stress tensor, $\bm{v}$ the velocity field and  $\bm{\rho}$ the density field of the material. Moreover, we let $\bm{b}$ denote the external body force per unit mass, e.g., that of gravity. The conservation of mass and momentum can then be written in conservative form as 
\begin{align}
    \label{mass_conservative}
    \der{\rho}{t}(\bm x, t)  + \nabla \cdot ( \rho \bm v ) (\bm x , t) &= 0\\
    \der{\rho \bm v}{t}(\bm x, t)  + \nabla \cdot ( \rho \bm v \otimes \bm v) (\bm x , t) &= \nabla \cdot \bm \sigma (\bm x, t) + \rho (\bm x, t) \bm b (\bm x, t), \quad \forall t>0, \; \bm{x} \in \Omega.
    \label{qdm_conservative}
\end{align}
where $\Omega$ is the domain where the solid is confined. 

\subsubsection{Depth integration on flat surface}
\label{flat}
We denote $h$ the height of the flow at any point, which can be seen as a function of the other coordinates $x, y$ and $t$. For any field $\phi : \mathbb{R}^3 \times \mathbb{R}_+ \longrightarrow E$ (where $E$ designates $\mathbb{R}$ or $\mathbb{R}^3$ ) we introduce the associated depth-averaged field $\overline{\phi}$ such as 
\begin{align*}
    \overline{\phi} : \big( x, y, t \big) \longmapsto \frac{1}{h(x, y, t)} \: \int_0^h \phi(x, y, z, t) dz 
\end{align*} 
The field $\phi'=\phi - \overline{\phi}$ is defined as the difference between the field and its depth-averaged field. This implies  $\displaystyle \int_0^h \phi' dz = 0$ and $\displaystyle \int_0^h \phi' \overline{\phi} \, dz = 0$. For each field, we also denote 
\begin{align}
    \|\phi\|_{\infty, \: h} (x, y, t) := \max_{z \in [0, h]} |\phi (x, y, z, t)|.
\end{align}
In order to simplify the notation, we also denote $\rbar{\phi}{z = z_0}$ for any real $z_0$, the function from $\mathbb{R}^2 \times \mathbb{R}_+$ to $E$ such that
$$
    \rbar{\phi}{z = z_0} (x, y, t) = \phi (x, y, z_0, t)
$$

\paragraph*{Assumptions}

The classic shallow-water assumptions are

\begin{itemize}
    \item The flow depth varies gradually and is small compared to the other dimensions of the flow. This is formalized by $ \varepsilon = \frac{h_0}{L} \ll 1$ where $h_0$ is the standard height of the flow and $L$ the characteristic length of the release zone.
    \item The material is incompressible: the density $\rho$ does not depend on position neither time.
    \item The flow surface is stress-free, i.e., the Cauchy stress tensor on the boundary is $\rbar{\bm{\sigma}}{z=h}=0$
    \item Velocity fluctuations within the depths are small compared to the average velocity which reads
    \begin{align}
    \frac{\|v'_x\|_{\infty, \: h}}{\overline{v_x}} \sim \frac{\|v'_y\|_{\infty, \: h}}{\overline{v_y}} = \mathcal{O}(\varepsilon).
    \label{ditribution_z}
    \end{align}
    \item The vertical velocity is small compared to the other velocity components i.e. \begin{align}
    \frac{\|v_z\|_{\infty, \: h}}{\overline{v_x}} \sim \frac{\|v_z\|_{\infty, \: h}}{\overline{v_y}} = \mathcal{O}(\varepsilon).
    \label{velocity_z}
\end{align}

\end{itemize} 

Under the hypothesis of incompressibility, the equation of mass conservation is therefore simplified as follows   
\begin{align}
    \label{3d_conservation_mass_inc}
    \nabla \cdot \bm{v} &= 0 \\
    \intertext{and the momentum conservation equation can thus be rewritten as}
    \label{3d_conservation_momentum_inc}
    \frac{\partial \bm{v}}{\partial t} + \nabla \cdot \big( \bm{v}\otimes \bm{v}\big) &= \frac{1}{\rho} \nabla \cdot \bm{\sigma} + \bm{b}
\end{align}

\paragraph*{Boundary conditions}

We have the following boundary conditions for the flow.
\begin{align}
    \label{Boundary_condition_h}
    v_z(z=h) &= \frac{\partial h}{\partial t} + v_x(z=h) \frac{\partial h}{\partial x} + v_y(z=h) \frac{\partial h}{\partial y} \\
    \label{Boundary_condition_0}
    v_z(z=0) &= 0.
\end{align}
where the first one expresses the kinematic condition on $z=h$ and the second expresses the non-porosity of the surface $z=0$.

\paragraph*{Depth-averaged mass conservation}

We integrate Eq.~\eqref{3d_conservation_mass_inc} for $z \in [0, h]$. We get
\begin{equation*}
    \int_0^h \nabla \cdot \bm{v} dz = 0,
\end{equation*}
which can be rewritten as
\begin{align*}
    \int_0^h \frac{\partial v_x}{ \partial x} dz + \int_0^h \frac{\partial v_y}{ \partial y} dz + \int_0^h \frac{\partial v_z}{ \partial z} dz = 0
\end{align*}
Applying Leibniz integration rule to each term, we obtain
\begin{align*}
    \frac{\partial (h\overline{v_x})}{\partial x} - v_x(z=h) \frac{\partial h}{\partial x} +
    \frac{\partial (h\overline{v_y})}{\partial y} -  v_y(z=h) \frac{\partial h}{\partial y} + v_z(z=h) &= 0
\end{align*}
Combined with the boundary condition described in Eq.~\eqref{Boundary_condition_h}, this leads to 
\begin{align}
    \frac{\partial h}{\partial t} + \frac{\partial (h \overline{v_x} )}{\partial x} + \frac{\partial (h \overline{v_y} )}{\partial y} = 0,
    \label{mass_cont_inc}
\end{align}
which is the depth-averaged equation of mass conservation.

\paragraph*{Depth-averaged momentum conservation}

With the same method, we integrate in the $z$-direction Eq.~\eqref{3d_conservation_momentum_inc} and get

\begin{align*}
    \rho h\left(\frac{\partial \overline{v}_x}{\partial t} +\overline{v}_x \frac{\partial \overline{v}_x}{\partial x} +\overline{v}_y \frac{\partial \overline{v}_x}{\partial y}\right) &= \frac{\partial (h \overline{\sigma}_{xx})}{\partial x} + \frac{\partial  (h \overline{\sigma}_{xy})}{\partial y} - \tau_{xz} + b_x \rho h. \\
    \rho h\left(\frac{\partial \overline{v}_y}{\partial t} +\overline{v}_y \frac{\partial \overline{v}_y}{\partial y} +\overline{v}_x \frac{\partial \overline{v}_y}{\partial x}\right) &=\frac{\partial (h \overline{\sigma}_{yy})}{\partial y} + \frac{\partial  (h \overline{\sigma}_{xy})}{\partial x} - \tau_{yz} + b_y \rho h.
\end{align*}
where $\tau_{xz} := \rbar{\sigma_{xz}}{z=0}$ and $\tau_{yz} := \rbar{\sigma_{yz}}{z=0}$. We define the depth-averaged material derivative operator for any depth-averaged field $\overline{\phi}$ as 
\begin{align*}
    \damatder{\overline{\phi}} = \der{\overline{\phi}}{t} + \olv_x \der{\overline{\phi}}{x} + \olv_y \der{\overline{\phi}}{y}.
\end{align*}
We can now write the non-conservative form of the momentum conservation:
\begin{align}
    \rho h \damatder{\olv_x} &= \frac{\partial (h \overline{\sigma}_{xx})}{\partial x} + \frac{\partial  (h \overline{\sigma}_{xy})}{\partial y} - \tau_{xz} + b_x \rho h.  \label{non_conservative_x_inc}\\
    \rho h \damatder{\olv_y} &=\frac{\partial (h \overline{\sigma}_{yy})}{\partial y} + \frac{\partial  (h \overline{\sigma}_{xy})}{\partial x} - \tau_{yz} + b_y \rho h.
    \label{non_conservative_y_inc}
\end{align}
Note that $\damatder{\olv_x}$ and $\damatder{\olv_y}$ are the components of the material acceleration $\overline{a}$. We can combine Eq.~\eqref{non_conservative_x_inc} and~\eqref{non_conservative_y_inc} in the vectorial equation 
\begin{equation}
    \overline{a} = \overline{\nabla} \cdot \big( h \ols \big) -  \bm \tau_{b} + \bm b \rho h.
    \label{non_conservative}
\end{equation}
where
\begin{align*}
\overline{a} = \begin{pmatrix}
\damatder{\olv_x} \\
\damatder{\olv_y} 
\end{pmatrix}, \quad \overline{\nabla} \cdot \big( h \ols \big) = \begin{pmatrix}
\frac{\partial (h \overline{\sigma}_{xx})}{\partial x} + \frac{\partial  (h \overline{\sigma}_{xy})}{\partial y} \\
\frac{\partial (h \overline{\sigma}_{yy})}{\partial y} + \frac{\partial  (h \overline{\sigma}_{xy})}{\partial x}  
\end{pmatrix}, \text{ and } \bm \tau_b = \begin{pmatrix}
\tau_{xz} \\
\tau_{yz} 
\end{pmatrix}.
\end{align*}

\subsubsection{Depth integration on complex topography}

In the previous integration, we supposed that the terrain was planar. 
This will be the case in most of our validation framework and applications of the model. Yet, some simulations on complex terrain will be presented which require a change of coordinate from Cartesian to curvilinear, which adds surface gradient terms to the depth-averaged momentum conservation equation. 
For the sake of clarity, the related  mathematical framework based on the work of~\cite{Bouchut2004} and~\cite{Vila2008} is presented in~\ref{complex_terr}. 
Note that this change of coordinate system, instead of a simple vertical projection, is necessary to recover multi-directional wave and crack propagation features.

\subsubsection{Constitutive model for snow slab avalanche}
\label{slab_mod}

The system composed of Eqs~\eqref{mass_cont_inc}, \eqref{non_conservative_x_inc} and ~\eqref{non_conservative_y_inc} presented in the last section is not closed. To close it, we have to choose a model relating the stress tensor~$\bm \sigma$ to the deformation. 
In this framework, we choose an elasto-plastic model, where the slab behaves elastically until a yield stress is reached, marking the onset of permanent deformations. 

\paragraph{Small strain tensor}

We denote the Lagrangian coordinate $\bm X$, i.e., the coordinate of the undeformed material. 
The map of deformation $\phi$ is then defined as
\begin{equation*}
    \bm u (\bm X, t) = \phi( \bm X, t) - \bm X
\end{equation*}
where $\bm u$ is the displacement of the material initially in position $\bm X$ between time $t$ and time $0$. The deformation gradient $\bm F$ is defined as the gradient of the deformation map, $ \bm{F} = \nabla \phi (\bm X, t)$. Note that the deformation gradient is related to the displacement by $\bm F = \nabla \bm u + \bm{I}$, where $\bm{I}$ is the identity tensor. The Green-Lagrange stain tensor $\bm E$ is defined as
\begin{equation*}
    \bm E = \frac{1}{2} \big( \bm F \bm F^T - \bm{I} \big)
\end{equation*}
which, using the expression for the deformation gradient, becomes
\begin{align*}
    \bm E = \frac{1}{2} \Big( \nabla \bm u + (\nabla \bm u)^T + \nabla \bm u \cdot (\nabla \bm u)^
    T \Big).
\end{align*}
We make the hypothesis that the displacement pf the slab is small, we supposed then that $\| \nabla \bm u \| \ll 1$. In this condition, the Green-Lagrange strain tensor can be approximated by the small deformation strain tensor 
\begin{equation}
    \bm \varepsilon = \frac{1}{2} \big( \nabla \bm u + \nabla \bm u^T \big).
    \label{strain_def}
\end{equation}

\paragraph{Elasto-plastic model for the slab}

We choose for the slab an additive elasto-plastic model,
\begin{equation*}
    \bm{\varepsilon} = \bm{\varepsilon_p} + \bm{\varepsilon_e}
    \label{elastic}
\end{equation*}
with $\bm \varepsilon_e$ satisfying a linear elastic law,
\begin{equation}
    \bm \sigma = \bm{\mathcal{D}}:\bm{\varepsilon_e}
\end{equation}
With $\bm{\mathcal{D}}$ is the fourth order elastic tensor depending of the material elastic properties (Young's modulus and Poisson's ratio). From the last equation we have  
\begin{equation}
    \bm \sigma = \bm{\mathcal{D}}:(\bm{\varepsilon} - \bm{\varepsilon_p}).
\end{equation}
In order to define the onset of plastic, irreversible, deformations, a yield criterion on the stress can be introduced. 
Here, we choose the Cohesive Cam Clay (CCC) criterion based on Critical State Soil Mechanics \citep{Roscoe1968}. 
This model was previously proposed to model snow mechanical behavior~\citep{Meschke1996,Gaume2018,Trottet2022}. 
In the space of the stress invariants $p$, the mean stress, and $q$, the von Mises equivalent stress, this criterion can be expressed as
\begin{equation}
   q(\bm \sigma)^2(1 + 2 \beta) + M^2 \big( p(\bm \sigma) + \beta p_0) \big) \big( p(\bm \sigma) - p_0) \big) \leq 0
    \label{MCC}
\end{equation}
where $p_0$ corresponds to the consolidation pressure and directly influences the size of the yield surface; $M$ is the slope of the cohesionless critical state line that controls the amount of friction inside the material and the shape of the yield surface; $\beta$ is the cohesion parameter that quantifies the ratio between tensile and compressive strengths. 
This yield surface is illustrated in the space of $p$ and $q$ in~Fig.~\ref{constitutive_laws}a. For all this framework, we consider the slab as purely brittle. Therefore, if the yield criterion is not respected, we consider the slab as broken and the stress tensor is set to 0.

In the depth-averaged framework, the shear in the $x-z$ and $y-z$ plane can be neglected compared to the other stress tensor component leading to a \emph{plane stress} hypothesis. Therefore the stress tensor used to computed the elastic and plastic deformation is the $2\times2$ matrix 
\begin{equation*}
    \bm \sigma = \begin{pmatrix}
\sigma_{xx} & \sigma_{xy}\\
\sigma_{xy} & \sigma_{yy} 
\end{pmatrix}.
\end{equation*}
The stress invariant are then computed as 
\begin{align}
    p(\bm \sigma) = -\frac{1}{2} \big(\sigma_{xx} + \sigma_{yy}\big), \quad \bm s (\bm \sigma) = \bm \sigma + p(\bm \sigma) Id, \quad q(\bm \sigma) = \sqrt{\frac{3}{2} \bm s : \bm s}.
\end{align}
Note that for the uniaxial tension and uniaxial compression the value of $p$ and $q$ are different than if computed with the full $3\times3$ stress tensor. 
This implies the need of transforming the CCC parameters $(p_0, M, \beta)$ to have the same tensile strength, compressive strength and shear resistance in both the full 3D model and the depth-averaged model.

\subsubsection{Weak layer modelling}

The weak layer - slab interaction will be modeled by a basal force $\tau_b$. The weak layer is considered as elastic-purely brittle.

\begin{figure}[H]
\centering
        \includegraphics[width=\linewidth]{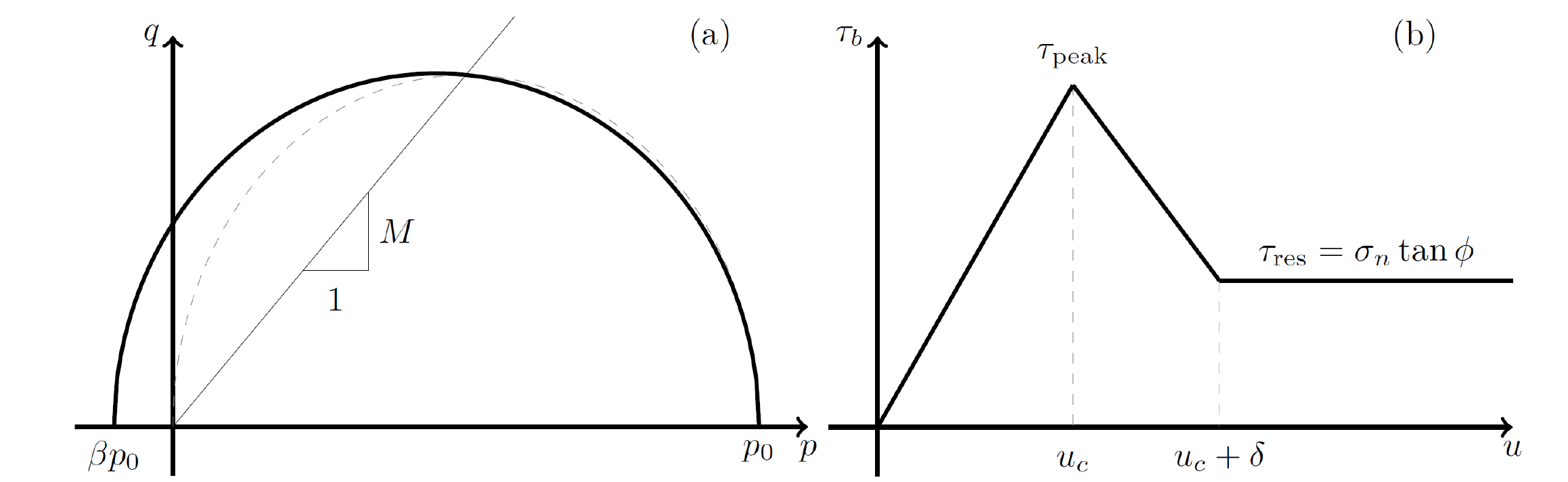}
     \caption{(a) Cohesive Cam Clay yield surface in the $p-q$ space. The dashed curve represents the cohesionless case. (b) Basal shear stress as a function of the displacement $u$ of the slab.}
     \label{constitutive_laws}
\end{figure}
In Fig.~\ref{constitutive_laws}b, the basal shear force is plotted as a function of the slab displacement norm $\bm u$. As long as the displacement is lower than the critical displacement, the basal stress $\tau_b$ increases with the displacement until it reaching the peak stress $\tau_{\text{peak}}$. Then, in order to model the quasi-brittle failure of the weak layer, the stress decreases down to the residual stress $\tau_{\text{res}} = \sigma_n \tan \phi$ with $\sigma_n$ the normal stress at the base of the material and $\phi$ the basal frictional angle between the slab and the weak layer. This represents the Coulomb model for the dry friction when the sliding of the slab on the weak layer occurs. The softening phase is associated with a plastic displacement $\delta$.\\

In the domain $u < u_c$ the model acts as a spring between the slab and the weak layer whose stiffness $k_{w}$ is the slope of the linear part. $k_w$ can be related to the geometrical and mechanical characteristics of the weak layer as 
$$
k_{w} = \frac{G_{w}}{D_{w}},
$$
 where $G_w$ is the shear modulus and $D_w$ is its thickness.
 We can thus give the expression of $\tau_b$ as a function of $u$:
 
\begin{equation}
   \tau_b (u) = \begin{cases}
        k_w \, u, \quad &\text{if $u<u_c$} \\
        \tau_{\text{res}}, \quad & \text{if $u>u_c + \delta$} \\
\end{cases} 
\label{cubic spline}
\end{equation}
 We can also define the cohesion $C_w$  between the slab layer as  
 $$
    C_w = \tau_{\text{peak}} - \tau_{\text{res}} = \tau_{\text{peak}} - \sigma_n \tan \phi.
 $$
As  $u_c = \tau_{\text{peak}} / k_w$, the weak layer model requires three parameters; the peak stress $\tau_{\text{peak}}$, thickness $D_w$ and  shear modulus $G_w$.

\subsection{Numerical model: DAMPM}

\subsubsection{Space and time discretization}

\begin{figure}
    \centering
    \includegraphics[trim={0 3.5cm 0 0},width=\linewidth]{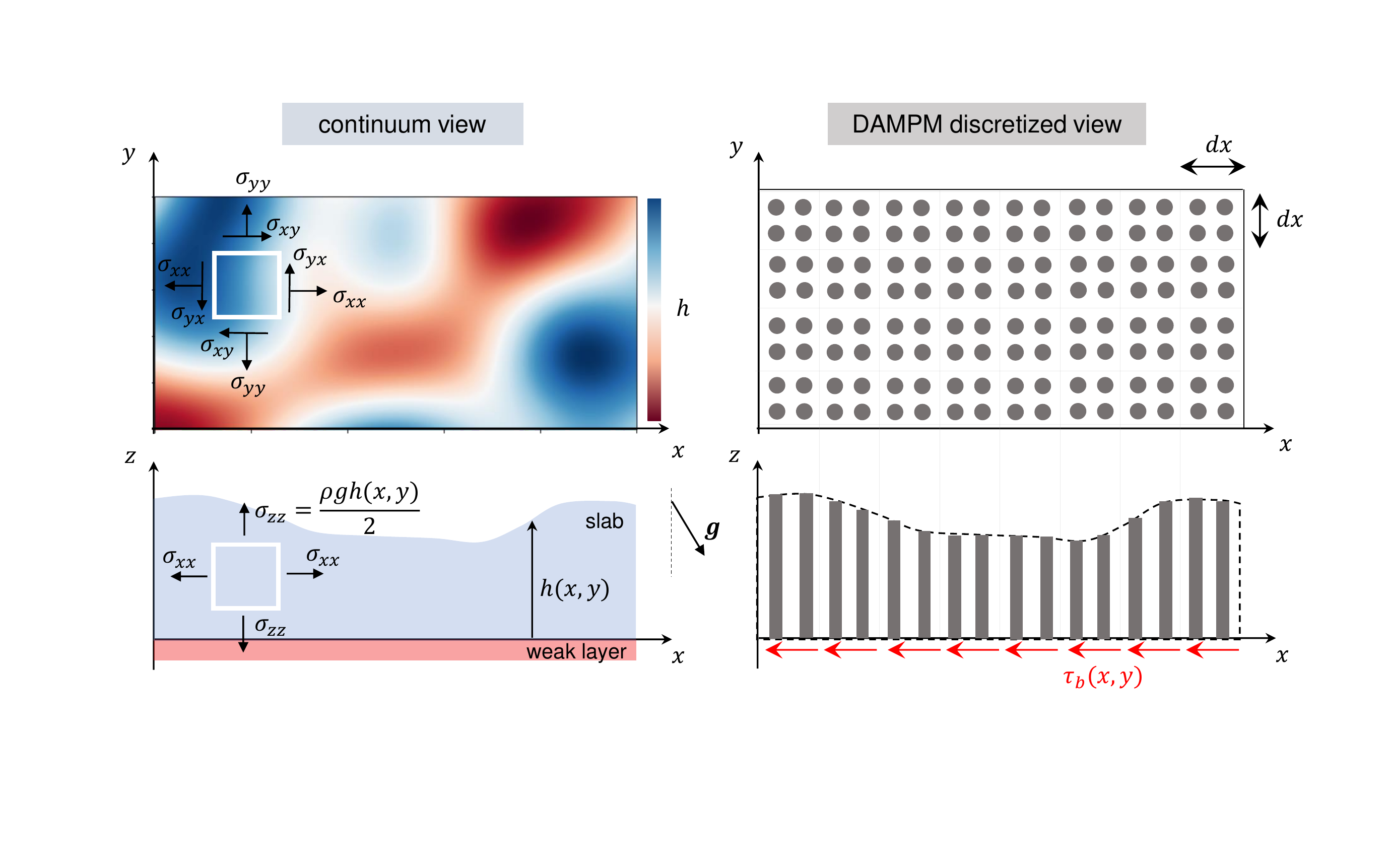}
    \caption{Illustration of some model assumptions and DAMPM discretization.}
    \label{DAMPM}
\end{figure}

In this section we develop the depth-averaged version of the MPM algorithm used to solve Eqs.~\eqref{non_conservative_x_inc}~and~\eqref{non_conservative_y_inc}. 
A sketch of the scheme is illustrated in Fig.~\ref{DAMPM}. 
A depth-averaged version of MPM has been introduced by \citep{Abe16} to simulate shallow debris flows. 
The principle of the method is very close to the classical MPM. 
We discretize the material domain with Lagrangian points, which unlike in classical MPM, are now "columns" with a certain variable height.
In addition to their height, these columns store information of velocity, stress and plastic deformation.
Furthermore, a background Eulerian grid is used to facilitate uncomplicated spatial differentiation. 
An interpolation scheme is used to transfer quantities between columns and grid nodes.

In the same spirit of the Finite Element Method (FEM), we seek a weak form of the depth-averaged equations.
To this end, we multiply Eqs.~\eqref{non_conservative_x_inc} and \eqref{non_conservative_y_inc} by two test function $w_x$ and $w_y$ and then integrate over the domain $\Omega$. Note that the rigorous formalism of which functional space the later functions belong will not be developed in this work. We use the same notation as in \citep{Abe16} with $d\Omega$ being the infinitesimal element of integration.
\begin{align*}
\int_{\Omega} \rho h \damatder{\olv_x} w_x d\Omega 
&= 
\int_{\Omega} \Big[ \der{h \ols_{xx}}{x} + \der{h \ols_{xy}}{y} \Big] w_x d\Omega - \int_{\Omega} \tau_{xz} w_x d\Omega + \int_{\Omega} b_x \rho h w_x d\Omega  \\
\int_{\Omega} \rho h \damatder{\olv_y} w_y d\Omega 
&= 
\int_{\Omega} \Big[ \der{h \ols_{xy}}{x} + \der{h \ols_{yy}}{y} \Big] w_y d\Omega - \int_{\Omega} \tau_{yz} w_y d\Omega + \int_{\Omega} b_y \rho h w_y d\Omega 
\end{align*}

Using Green's theorem we have 
\begin{align*}
\int_{\Omega} \rho h \damatder{\olv_x} w_x d\Omega &= - \int_{\Omega} h \ols_{xx} \der{w_x}{x} d\Omega - \int_{\Omega} h \ols_{xy} \der{w_x}{y} d\Omega - \int_{\Omega} \tau_{xz} w_x d\Omega + \int_{\Omega} b_x \rho h w_x d\Omega \\
\int_{\Omega} \rho h \damatder{\olv_y} w_y d\Omega &= - \int_{\Omega} h \ols_{xy} \der{w_y}{x} d \Omega - \int_{\Omega} h \ols_{yy} \der{w_y}{y} d\Omega - \int_{\Omega} \tau_{yz} w_y d\Omega + \int_{\Omega} b_y \rho h w_y d\Omega 
\end{align*}

We define then $\bm \tau_b^S = \frac{1}{h} \bm \tau_b$ such that 
\begin{align}
\int_{\Omega} \rho h \damatder{\olv_x} w_x   d\Omega &= - \int_{\Omega} h \ols_{xx} \der{w_x}{x} d\Omega - \int_{\Omega} h \ols_{xy} \der{w_x}{y} d\Omega - \int_{\Omega} h \tau_{xz}^S w_x d\Omega + \int_{\Omega} b_x \rho h w_x d\Omega \label{weak_formula_x} \\
\int_{\Omega} \rho h \damatder{\olv_y} w_y d\Omega &= - \int_{\Omega} h \ols_{xy} \der{w_y}{x} d \Omega - \int_{\Omega} h \ols_{yy} \der{w_y}{y} d\Omega - \int_{\Omega} h \tau_{yz}^S w_y d\Omega + \int_{\Omega} b_y \rho h w_y d\Omega \label{weak_formula_y}
\end{align}

We decompose our material in $n_p$ columns. these columns will move with the material during the simulation. As in standard MPM, we assume that the mass is concentrated at each column. We have then 
\begin{equation}
    \forall \bm \in \Omega, \; t \in \mathbb{R}_+, \quad \rho(\bm x, t) h(\bm x, t) = \sum_{p=1}^{n_p} m_p \; \delta (\bm x - \bm x_p)
    \label{discret_p}
\end{equation}
where $\delta$ is the Dirac distribution with dimension of inverse of a surface and $\bm{x}_p$ is the position of the column~$p$. Note that $\bm{x}_p$ is a function of time. Here, because we assume that the material is incompressible, this simplifies to
\begin{equation}
    \forall \bm \in \Omega, \; t \in \mathbb{R}_+, \quad  h(\bm x, t) = \sum_{p=1}^{N_p} V_p \; \delta (\bm x - \bm x_p), 
    \label{discret_p_}
\end{equation}
 with $V_p$ is the constant volume of the column. 
 
 Until the end of this subsection, we present the discretization only for Eq.~\eqref{weak_formula_x}, however, the discretization for Eq.~\eqref{weak_formula_y} is analogous.
 Inserting Eq.~\eqref{discret_p_} into Eq.~\eqref{weak_formula_x}, we have
\begin{align*}
    \sum_p \rho V_p \; \ola_{x, p} \; w_{x, p}  = &- \sum_p V_p \ols_{xx, p} \der{w_{x, p}}{x} - \sum_p V_p \ols_{xy, p} \der{w_{x, p}}{y} \\ 
    &- \sum_p V_p \tau^S_{xz, p} w_{x, p} + \sum_p V_p b_{x, p} \rho w_{x, p}
\end{align*}
where again the subscript $p$ of a quantity correspond to its evaluation in $\bm x_p$. 
We can apply this discretization at any time $t^k$, denoting for any function $\phi$,  $\phi^k = \phi(t^k)$, the previous equation gives
\begin{align}
    \sum_p \rho V_p \; \ola_{x, p}^k \; w_{x, p} = &- \sum_p V_p \ols_{xx, p}^k \der{w_{x, p}}{x} - \sum_p V_p \ols_{xy, p}^k \der{w_{x, p}}{y} \\ 
    &- \sum_p A_p \tau^k_{xz, p} w_{x, p} + \sum_p \rho V_p b_{x, p}^k \; w_{x, p}
    \label{sum_over_p}
\end{align}
where $A_p = V_p / h$ is the area of the column $p$. 
We want to perform the computation on the grid node. To this end, we introduce interpolation functions $N_i(\bm x)$ centred on node $i \in \{1,..., n_g\}$. 
We can then approximate
\begin{equation}
    \ola_{x, p}^k = \sumi \ola_{x, i}^k \; N_{ip}^k, \quad w_{x, p} = \sumi w_{x, i} \; N_{ip}^k,
    \label{interpolation_grid}
\end{equation}
where $\ola_{x, i}^k = \ola_x(\bm x_i, t^k)$,  $N_{ip}^k = N_i\big(\bm x_p(t^k)\big)$ and $w_{x, i} = w_x(\bm x_i)$. The gradients can be interpolated as well with the equality
 \begin{equation*}
     \nabla w_{x, p} = \sumi w_{x, i} \; \nabla N_{ip}^k, \quad \nabla N_{ip}^k = \nabla N_i \big( \bm x_p(t^k) \big)
 \end{equation*}
 Note that the function $N$ must satisfy the following conditions \citep{Gao2017}:
 \begin{itemize}
     \item partition of unity: $\sum_i N_{ip}^k = 1$, $\forall p \in \{ 1, ..., n_p\}$
     \item non-negativity: $N_{ip}^k \geq 0$, $\forall p \in \{ 1, ..., n_p\}$ and $\forall i \in \{ 1, ..., n_g\}$
     \item derivability: $N_i \in \mathcal{C}^1$, $\forall i \in \{ 1, ..., n_g\}$
     \item interpolation: $\bm x_i = \sum_p N_{ip}^k x_p$
     \item local support: $N_{ip} > 0$ only if $\bm x_i$ and $\bm x_p$ are close enough
 \end{itemize}

 We can now express Eq.~\eqref{sum_over_p} with quantities over the grid
 \begin{align*}
    \sump \sumi \sumj \rho V_p \ola_{x, j}^k N_{jp}^k w_{x, i} N_{ip}^k  = &- \sump \sumi V_p  \ols_{xx, p}^k w_{x, i} \der{N_{ip}^k}{x} - \sump \sumi V_p  \ols_{xy, p}^k \der{N_{ip}^k}{y} w_{x, i} \\
    &- \sump \sumi A_p  \tau^k_{xz, p} w_{x, i} N_{ip}^k + \sump \sumi \rho V_p b_{x, p}^k \; w_{x, i} N_{ip}^k
 \end{align*}
Defining the mass matrix at time $k$ as
 \begin{equation*}
     m_{ij}^k = \sump m_p N_{ip}^k N_{jp}^k
 \end{equation*}
where $m^k_p = \rho V^k_p$ is the mass of the column $p$, the system becomes
\begin{align*}
    \sumi \sumj m_{ij}^k \ola_{x, j}^k w_{x, i} = &- \sumi \Bigg[ \sump \bigg[ V_p  \ols_{xx, p}^k \der{N_{ip}^k}{x} + V_p  \ols_{xy, p}^k \der{N_{ip}^k}{y} \bigg] \Bigg] w_{x, i} \\
    & - \sumi \Bigg[ \sump A_p \tau_{xz, p}^k N_{ip}^k \Bigg] w_{x, i} + \sumi \Bigg[ \sump \rho V_p b_{p, x} N_{ip}^k \Bigg] w_{x, i}.
\end{align*}

As the interpolation functions satisfy $\sumi N_{ip}^k = 1$, the mass matrix can be approximated by a symmetric positive diagonal matrix (the so-called lumped mass approach) \citep{JIANG}, such that
\begin{equation*}
    \sumj m_{ij}^k \ola_{x, j}^k \approx m_i^k \ola_{x, i}^k
\end{equation*}
where
\begin{equation*}
    m_i^k = \sumj m_{ij}^k
\end{equation*}
gives the mass at node $i$.

Defining the internal forces $f^{\text{int}, k}_{x, i}$ and the external forces $f^{\text{ext}, k}_{x, i}$ for every $i \in \{1, ..., n_g \}$  as
\begin{align}
    f^{\text{ext}, k}_{x, i} = &\sump \rho V_p b_{x, p} N_{ip}^k \label{forces1} \\ 
    f^{\text{int}, k}_{x, i} = - &\sump \bigg[ V_p  \ols_{xx, p}^k \der{N_{ip}^k}{x} + V_p  \ols_{xy, p}^k \der{N_{ip}^k}{y} + \tau_{xz, p}^k N_{ip}^k \bigg]
    \label{forces2}
\end{align}
the discretized equation becomes
\begin{equation*}
    \sumi m_{i}^k \ola_{x, i}^k w_{x, i} = \sumi \big( f^{\text{ext, k}}_{x, i} + f^{\text{int, k}}_{x, i}\big) w_{x, i}.
\end{equation*}

This relation being true for every sequences $\big( w_{x, i} \big)_{i \in \{ 1, ... ,n_g\} }$, we have the equality for each term.

\begin{equation}
   \quad m_{i}^k \ola_{x, i}^k = \big( f^{\text{ext, k}}_{x, i} + f^{\text{int, k}}_{x, i}\big)
   \text{\ \ \ } \forall \text{\ } i \in \{ 1, ... ,n_g\}.
    \label{newton}
\end{equation}

\subsubsection{Algorithm}
\label{meth_num}
Here, the algorithm for one time step of the Depth-Averaged Material Point Method is outlined. 
It relies on the elastic predictor - plastic corrector scheme of computational elastoplasticity, where a trial stress state is computed and projected back to the yield surface if permanent deformations occurred~\citep{souza}. 
A summary of the DAMPM algorithm is provided in Algorithm~\ref{algorithm}.
\begin{itemize}

    \item[1.]
    An Eulerian background grid with cell size $\Delta x$ is created to cover the material domain. If $t>0$, the grid is only extended or reduced to cover the updated material domain.
    \item[2.]
    The mass and velocity on the grid nodes at time $t^k$ is interpolated from the particles in what is called the Particles to Grid (P2G) step,
    $$ m_{\bm i}^k = \sump m_p N_{i, p}^k$$
    \begin{equation}
        \bm \olv_{\bm i}^k = \sump \bm \olv_p^k m_p N_{ip}^k / m_{\bm i}^k
        \label{P2G}
    \end{equation}
    Note the interpolation of the momentum divided by the lumped mass matrix component, guaranteeing conservation of momentum between material points and grid nodes.
    The interpolation functions used in this study are cubic B-splines,
    \begin{equation}
        N(u) = \left\{ \begin{array}{ll}
            \dfrac{1}{2} |u|^3 - |u|^2 + \frac{2}{3} & \text{if } |u| \leq 1  \\
            \dfrac{1}{6} (2 - |u|)^3 & \text{if } |u| \in [1, 2] \\
            0 & \text{if } |u| \geq 0 
        \end{array}
        \right.
    \end{equation}
    such that 
    $$
    N_{ip} = N_i(\bm x_p) = N\Big( \frac{x_i - x_p}{\Delta x}\Big) N \Big(\frac{y_i - y_p}{\Delta  y}\Big).
    $$
    
    \item[3.] 
    Based on the stress tensor $\bm \ols_p^k$ and the external force $\bm b_p^k$ on the particles, the forces $\bm f_{i}^{\text{int}, k}$ and $\bm f_{i}^{\text{ext}, k}$ on the grid nodes  are computed as described in Eqs.~\eqref{forces1} and~\eqref{forces2}.
    \item[4.] Using Eq.~\eqref{newton}, the acceleration on the grid nodes is obtained from the previously computed forces
    \begin{equation}
        \bm \ola_i^k = (\bm f_{i}^{\text{int}, k} + \bm f_{i}^{\text{ext}, k}) / m_{i}^k
        \label{accel}
    \end{equation}
    from which a grid velocity can be computed
    \begin{equation}
        \bm \olv_{\bm i}^{L} = \bm \olv_{\bm i}^{k} + \Delta t \bm \ola_i^k
        \label{velocity_update}
    \end{equation}
    \item[5.] 
    Instead of immediately computing the increment of strain and stress we interpolate the velocity and the position on the particles, in what is called the Grid to Particles~(G2P) step,
    \begin{align}
        \label{G2P_velocity}
        \bm{\overline{v}}_p^{k+1} &= \bm{\overline{v}}_p^{k} + \sum_{i=1}^{n_g} \big( (\Delta t\bm{\overline{a}}_i^k) C_\text{flip} + \bm{\overline{v}}_i^L (1-C_\text{flip}) \big) N^k_{ip}   \\
        \label{G2P_position}
        \bm{x}_p^{k+1} &= \bm{x}_p^{k} + \Delta t \sum_{i=1}^{n_g} \bm{\overline{v}}_i^L N^k_{ip}
    \end{align}

\begin{algorithm}
    \caption{Depth-averaged material point method (DAMPM)}
\begin{algorithmic}
    \While{$t < T$} 
    \State 1. create/expand/reduce the Eulerian grid around the particle domain
    \State 2. P2G: interpolate particle velocity and mass to the grid with Eqs.~\eqref{P2G}
    \State 3. compute grid forces $\bm{f}^\text{int}_i$ and $\bm{f}^\text{ext}_i$ with Eqs.~\eqref{forces1}~and~\eqref{forces2} depending on $\bm\ols_p$, $\tau_{xz,p}$, $\tau_{yz,p}$ and $\bm{b}_{p}$
    \State 4. compute grid acceleration $\bm{a}_i$ with Eq.~\eqref{accel} and grid veloctiy $\bm{v}_i^L$ with Eq.~\eqref{velocity_update}
    \State 5. G2P: update the particle velocity $\bm{v}_p$ and position $\bm{x}_p$ with Eqs.~\eqref{G2P_velocity}~and~\eqref{G2P_position}
    \State 6. update the grid velocity $\bm{v}_i$ from particle velocity with Eq.~\eqref{MUSL} 
    \State 7. compute the strain tensor $\Delta \varepsilon_p^{\text{trial}}$ with Eq.~\eqref{strain_computation}
    \State 8. update the height $h_p$ with Eq.~\eqref{update_height}
    \State 9. compute the stress $\bm{\ols}_p^{\text{trial}}$ with Eq.~\eqref{sigma_trial}
    \State 10. compute the yield criterion $f\big(p(\bm{\ols}_p^{\text{trial}}), q(\bm{\ols}_p^{\text{trial}})\big)$ with Eq.~\eqref{MCC}
    \If{$f\big(p(\bm{\ols}_p^{\text{trial}}), q(\bm{\ols}_p^{\text{trial}})\big) \leq 0$}
        \State update the stress $\bm \ols_p$ with Eq.~\eqref{stress_update}
    \Else 
        \State set the stress $\bm \ols_p$ to $\bm 0$ 
    \EndIf
    \State 11. adapt $\Delta t$ according to with Eqs.~\eqref{CFL_speed}~and~\eqref{CFL_elastic} and update time $t = t+\Delta t$ 
    \EndWhile
\end{algorithmic}
\label{algorithm}
\end{algorithm}
    
    where $C_\text{flip}$ is the FLIP-PIC ratio  \citep{stomakhin}. $C_\text{flip} = 0$ corresponds to a pure PIC scheme where we obtain particle velocity as an interpolation from grid velocity directly. On the other hand, $C_\text{flip} = 1$ corresponds to a pure FLIP scheme where only the increment of grid velocity is interpolated, thus reducing dissipation at the cost of stability. In this work, we use $C_\text{flip} = 0.9$ as a balance between reducing dissipation and maintaining stability.
    \item[6.] 
    Following the so-called MUSL algorithm \citep{SULSKY1995}, the grid velocities are interpolated from the updated particle momentum before they are used to calculate the strain increment. 
    \begin{equation}
        \bm{\overline{v}}_i^{k+1} = \frac{1}{m_i^k}\sum_{p=1}^{n_p} m_p \bm{\overline{v}}_p^{k+1} N^k_{ip}
        \label{MUSL}
    \end{equation}
    While this step is optional, it improves stability.
    \item[7.]
    Relying on the grid velocities, Eq.~\eqref{strain_def} gives a trial strain increment on the particles,
    \begin{align}
    \Delta \varepsilon^\text{trial}_p = \frac{\Delta t}{2} \left( \sumi \nabla N^{k}_{ip} (\olv^{k+1}_i)^T + \sumi \big( \nabla N^{k}_{ip} \big)^T \olv^{k+1}_i \right)
    \label{strain_computation}
    \end{align}

    \item[8.]
    The height of each particle is updated according to the strain increment,
    \begin{align}
        h^{k+1}_p = \frac{h^k_p}{1 + \text{tr}(\Delta \varepsilon^\text{trial}_p)}
        \label{update_height}
    \end{align}
    
    \item[9.] The trial stress tensors can be computed depending on 
    \begin{equation}
        \bm \ols^\text{trial}_p = \bm{\mathcal{D}}^e :\bm \varepsilon^\text{trial}_p
        \label{sigma_trial}
    \end{equation}
    with $\bm{\mathcal{D}}^e$ the fourth order tensor in plane stress model.
    \item[10.] 
    If the trial stress tensor satisfies the yield criterion, the trial strain and stress tensors are taken as the new strain and stress tensors, respectively, i.e.,
    \begin{align}
        \label{strain_update}
        \bm \varepsilon^{k+1}_p &= \bm \varepsilon^\text{trial}_p  \\
        \bm \ols^{k+1}_p &= \bm \ols^\text{trial}_p
        \label{stress_update}
    \end{align}
    Otherwise, the stress vanishes as we assume the material to be purely brittle
    $$\bm \ols^{k+1}_p = \bm 0$$
    \item[11.]
    An adaptive time-stepping is used where the time step $\Delta t$ according to two conditions. The first is the Courant-Friedrich-Lewy (CFL) condition, resulting in
    \begin{equation}
        \Delta t \leq \frac{C_{\text{cfl}} \Delta x}{\max_{p = 1, \ldots, N_p } \left\lVert v_p \right\rVert_{L^2} }
        \label{CFL_speed}
    \end{equation}
    where $C_\text{cfl} < 1$.
    In addition, the time step $\Delta t$ must be smaller than the time an elastic wave takes to travel the distance $\Delta x$. The elastic wave speed is given by $\sqrt{E/\rho}$ thus the second condition imposes
    \begin{equation}
        \Delta t \leq C_\text{elastic} \frac{\Delta x}{\sqrt{E/\rho}}
        \label{CFL_elastic}
    \end{equation}
    where $C_\text{elastic} < 1$.
\end{itemize}

\section{Results}
\begin{figure}
    \centering
    \includegraphics[width=0.6\textwidth]{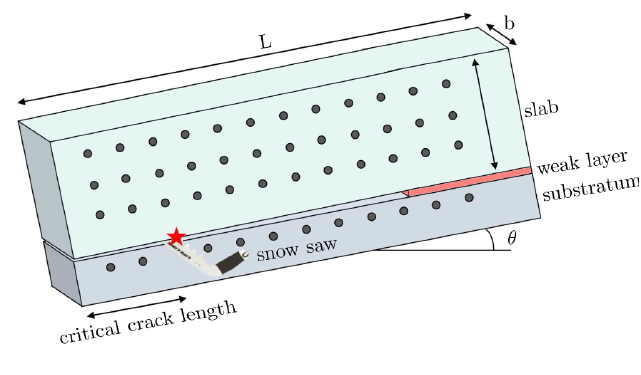}
    \caption{Schematic of the Propagation Saw Test (PST).}
    \label{PST_Schematic}
\end{figure}

In order to validate our model we perform simulations of the so-called Propagation Saw Test (PST). The PST is a classical snow fracture experiment used by avalanche researchers and practitioners to evaluate crack propagation propensity. It consists of isolating a column of snow and cutting through a previously identified weak layer, using a snow saw. This cutting procedure amounts to create a crack of increasing length until a so-called critical crack length is reached leading to self-sustained crack propagation. The PST setup is sketched in Fig. \ref{PST_Schematic}.

First, simulations with an elastic slab are made, where we can compare to analytical solutions for the critical crack length, the onset of crack propagation and the slab tensile failure distance. Second, simulations with a elastic-brittle slab are performed in which we evaluate the distance to the first slab fracture and compare to analytical relations. Third, we perform multi-directional fracture simulations to study crack propagation speeds in down/up slope and cross slope directions on a in a three dimensional slope configuration. Finally, we analyse slab fracture characteristics in mixed-mode crack propagation simulations.

Table~\ref{PST_elastic_table} shows parameters used throughout all simulations.

\begin{table}[h]
    \begin{adjustbox}{width=0.75\columnwidth,center}
    \rotatebox{0}{
    \begin{tabular}{llcccc}
         \hline
                     & & Case 1 & Case 2 & Case 3 & Case 4 \\
         \hline
            Geometry & Length $L_1$ ($\mathrm{m}$) & 40 & 40 & 50 & 40\\
                     & Width $L_2$ ($\mathrm{m}$) & 0.3 & 0.3 & 50 & 120 \\
                     & Slope angle $\theta$ ($\mathrm{^{\circ}}$) & & \multicolumn{2}{c}{45} & \\
         \hline
            Slab & Density $\rho^{init.}$ ($\mathrm{kg \cdot m^{-3}}$) & & \multicolumn{2}{c}{250} & \\
                    & Height $h$ ($\mathrm{m}$) & & \multicolumn{2}{c}{0.5} & \\
                     & Young's modulus $E$ ($\mathrm{MPa}$) & & \multicolumn{2}{c}{10} &\\
                     & Poisson's ratio $\nu$ & & \multicolumn{2}{c}{0.3} & \\
                     & Friction angle $\phi$ ($^\circ$) & & \multicolumn{2}{c}{27} & \\
             & $p_0$ ($\mathrm{kPa}$) & - & 30 & - & 15 \\
            & $M$ & - & 1.7 & - & 1.7 \\
            & $\beta$ & - & 0.1 & - & 0.1 \\
                     
         \hline
          Weak layer & Thickness $ D_w$ ($\mathrm{m}$) & & \multicolumn{2}{c}{0.125} &\\
                     & Young's modulus $E$ ($\mathrm{MPa}$) & & \multicolumn{2}{c}{1} &\\
                     & Poisson's ratio $\nu$ & & \multicolumn{2}{c}{0.3} & \\
                     & Shear strength $\tau_\text{peak}$ ($\mathrm{Pa}$) & & \multicolumn{2}{c}{1450} & \\
         \hline
            Numerical & Cell size $\Delta x$ ($\mathrm{cm}$) & 2 & 2 & 10 & 10 \\
                     & CFL constraint $C_\text{cfl}$ & & \multicolumn{2}{c}{0.5} & \\
                     & Elastic wave constraint $C_\text{elastic}$ & & \multicolumn{2}{c}{0.5} & \\
                     & FLIP/PIC ratio $C_\text{flip}$ & & \multicolumn{2}{c}{0.9} & \\
         \hline
    \end{tabular}
    }
    \end{adjustbox}
    \caption{Table of parameters for the different simulations presented in this work. The values in the middle of the table are valid for all the cases.}
    \label{PST_elastic_table}
\end{table}


\begin{figure}
    \includegraphics[width=0.9\textwidth]{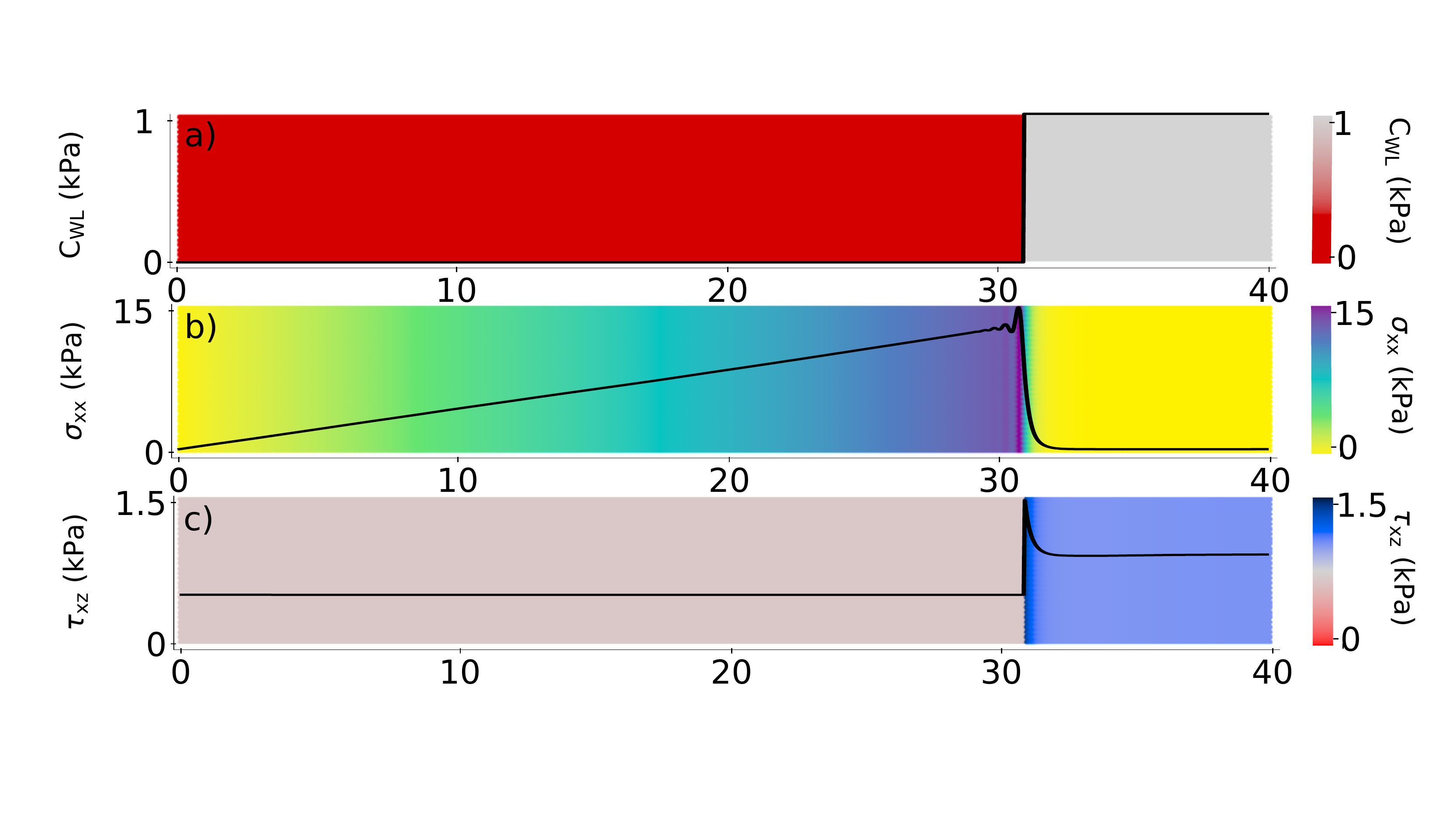}
    \caption{Top view of a PST simulation. From top to bottom, we represented the cohesion, the longitudinal stress and the basal stress}
    \label{PST_Elastic}
\end{figure}

\subsection{PST crack propagation with an elastic slab}

Fig.~\ref{PST_Elastic} show the cohesion (a), the tensile stress (b) and the shear stress in the weak layer (c) in a PST simulation during dynamic crack propagation at time ($t^*=t-t_c= 0.25$ s, $t_c:=$ time corresponding the onset of crack propagation). The parameters for this simulation correspond to case 1 of Table~\ref{PST_elastic_table}. The crack tip is identified based on cohesion which shows a transition between the initial value of $\approx$ 1kPa, and zero (Fig.~\ref{PST_Elastic}a). This transition is very sharp because the plastic softening displacement $\delta$ was set to zero. This transition in cohesion is directly linked to the shear stress in the weak layer $\tau_{xz}$ which reaches the weak layer strength $\tau_{p}=1.45$kPa (Fig.~\ref{PST_Elastic}b). The shear stress is characterised by three zones: far from the crack tip in the undisturbed region, the shear stress is equal to the stress induced by the projected slab weight $\tau_g=\rho g h \sin\theta$. Behind the crack tip, cohesion is zero and thus the shear stress is equal to the residual frictional stress. In between these two regions, the shear stress decreases exponentially from its maximum value $\tau_p$ to $\tau_g$ with a characteristic length $\Lambda$ (see below). In addition, the tensile stress in the slab increase linearly from the origin to the crack tip, where it is maximum. It then decreases exponentially to zero ahead of the crack. We note some small oscillation behind the crack tip which are very likely due to the limited amount of damping in our simulations.

In a spirit of verification, we performed several simulations with different values of the slope angle and computed the (super) critical crack length corresponding to the onset of crack propagation. We recall that due the depth-averaged nature of the model, we do not simulate the anticrack propagation regime but only the supershear mode of crack propagation as described in \cite{Trottet2022}. We thus refer to critical lengths as super critical in the later. In our particular case (PST) and with the assumption presented above, an analytical solution can be found \citep{gaume2013} and is given below:
\begin{equation}
    a_{sc} = \Lambda \bigg( 1 - \frac{\tau_p}{\tau_g} \bigg) \frac{\tan(\theta)}{\tan(\theta) - \tan(\phi)}
    \label{critical_length}
\end{equation}

\begin{figure}[H]
    \centering
    \includegraphics[width=\textwidth]{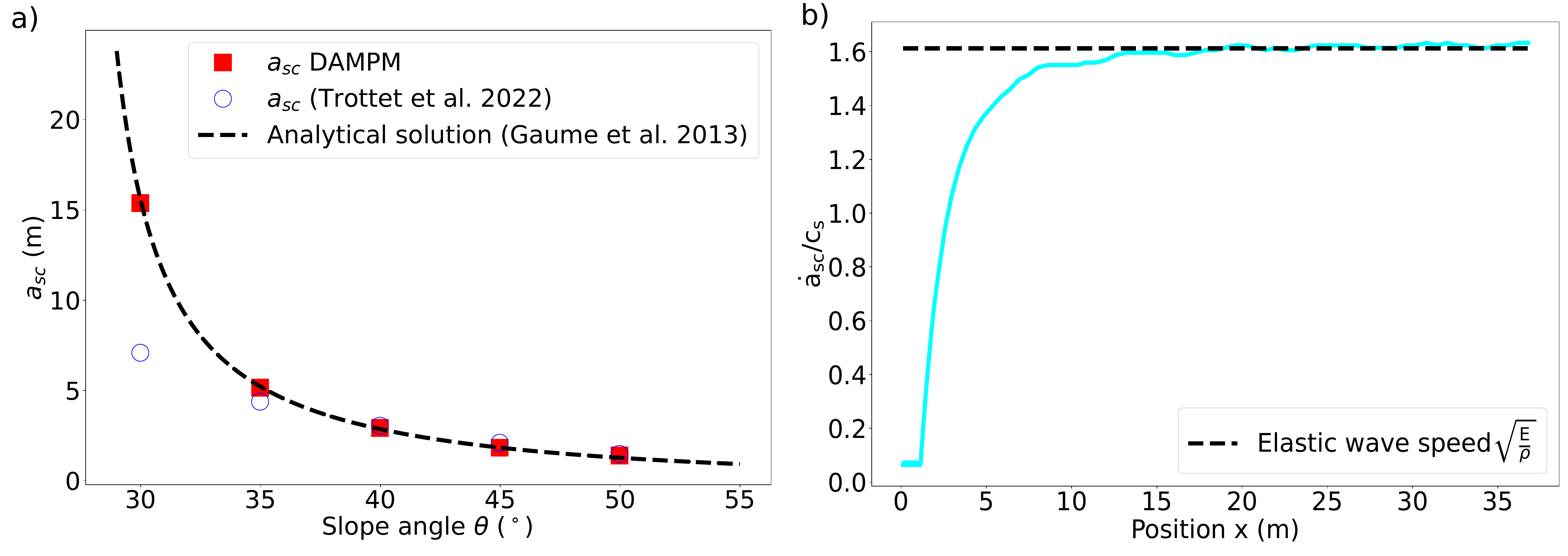}
    \caption{Left: comparison between the simulated super critical length using DAMPM, a 3D model \citep{Trottet2022} and the analytical solution in quasi-static regime for different angle. Right: Crack velocity profile as a function of the position.}
    \label{crack_speed}
\end{figure}

Fig.~\ref{crack_speed}a shows the supercritical crack length $a_{sc}$ as a function of slope angle for 3D simulations \citep{Trottet2022}, depth-averaged simulations (this study) and the analytical solution in Eq.~\eqref{critical_length}. We report an excellent agreement between our model and the expected solution. In fact the agreement is better than for 3D simulations, which was expected. Indeed, in 3D simulation, the collapse amplitude leads to a reduction in the effective friction coefficient of the weak layer and thus a reduction of $a_{sc}$ close to the frictional limit.

Another verification can be made by analysing the crack propagation speed $\dot{a}_{sc}$ limit (Fig.~\ref{crack_speed}b). We observe a strong increase in propagation speed between $a_{sc}$ and a distance of around 10 m where the speed levels off at an asymptotic value close to 1.6 $c_s=\sqrt{E/\rho}$. This is in line with the results of \citet{Trottet2022} and with expectations regarding the form of the 1-D equations which suggest that we should converge to the longitudinal wave speed.

\subsection{PST crack propagation with an elastic-brittle slab}

\begin{figure}
    \centering
    \includegraphics[width=\textwidth]{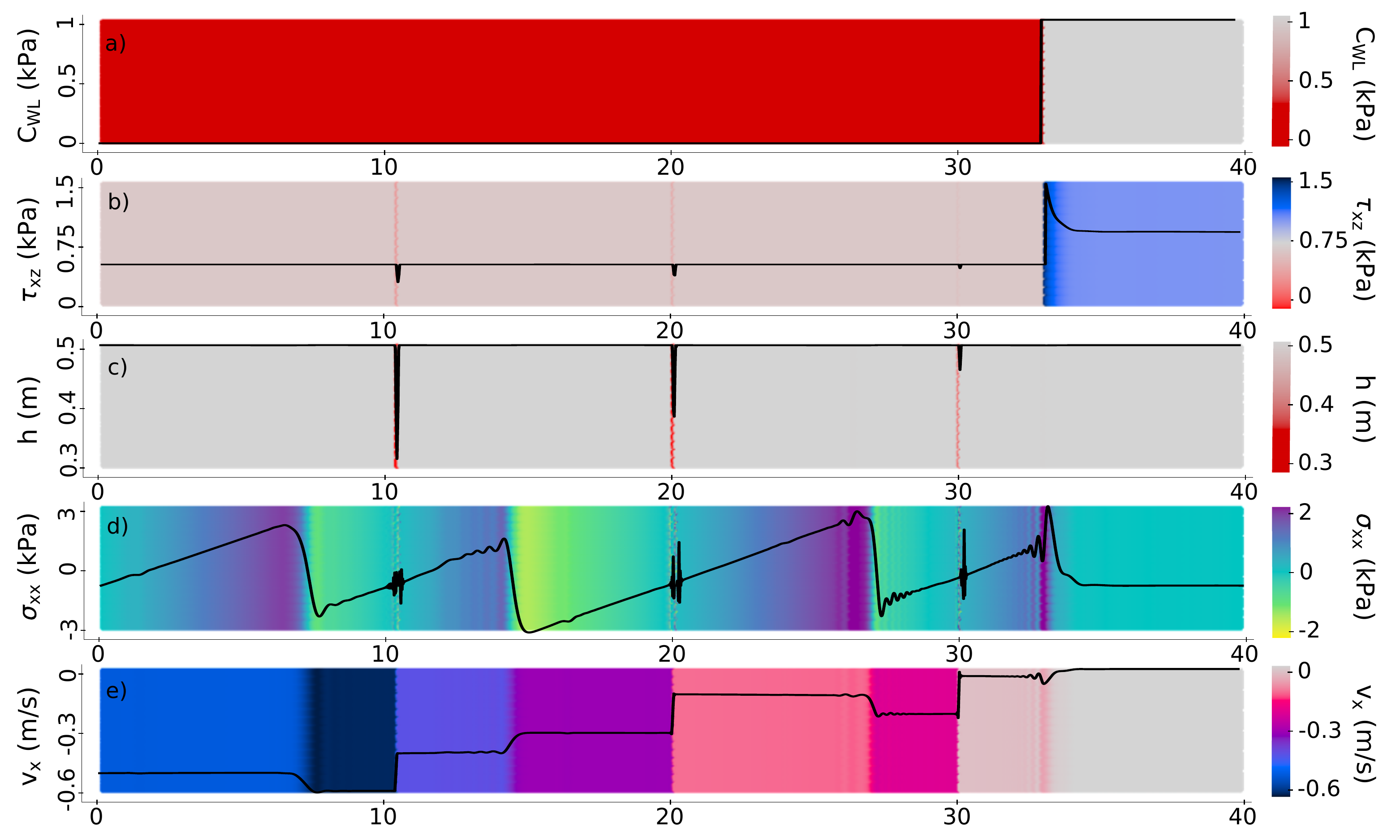}
    \caption{Top view of a PST simulation in which tensile fracture of the slab is enabled. From top to bottom, we represented the cohesion, the basal stress, the height, the longitudinal stress and the tangential velocity.}
    \label{PST_tensile}
\end{figure}

In this section, we relax the assumption of a purely elastic slab and perform the simulation with a slab failure criterion (see Section \ref{slab_mod}). Fig.~\ref{PST_tensile} shows the cohesion (a), the basal stress (b) the height (c), the longitudinal stress (d) and finally the velocity (e) as a function of the position $x$ of such a simulation at time $t^* = 0.3$s. The parameters used for the simulation are the same as the elastic simulation (see case 2 of Table \ref{PST_elastic_table}) and with a tensile strength $\sigma_t = 7$kPa. 

Tensile failures are characterised by a decreasing peak in the height profile (Fig.~\ref{PST_tensile}c). In this simulation, we observe 3 different failures at position $x \sim 10$m, $x \sim 20$m and $x \sim 30$m. As the fracture gap is growing with time, the height decrease is more pronounced in the first fracture. At each fracture interface, the slab stress is set to zero. As there is almost no damping in our simulations, we have small spatial oscillations in the tensile stress profile. In addition and for the same reason, the tensile stress is oscillating in time between compression and tension between two fractures, oscillations which do not attenuate in time (this can only be seen by watching the temporal evolution of these profiles in Supplementary video 2). At each failure, we observe a discontinuity of the down-slope velocity. The value of the velocity between two fractures is oscillating in time around a plateau in relation to the longitudinal stress oscillation described above. As the residual friction is computed as $\tau_r = \rho g h \tan(\phi)$, the decreasing peak in the height profile at each fracture will also be observed in the basal forces (Fig.~\ref{PST_tensile} b). The tensile fractures in the slab do not stop the propagation of the crack in the weak layer. The fractures have also very little impact on the profile of the stress and the basal force after the crack tip, which are the same as in the simulations with a purely elastic slab.  Again, the discontinuity of the basal force profile at the crack tip is due to $\delta$, the softening length of the weak layer which is equal to zero in the presented simulations.

\begin{figure}
    \includegraphics[width=\textwidth]{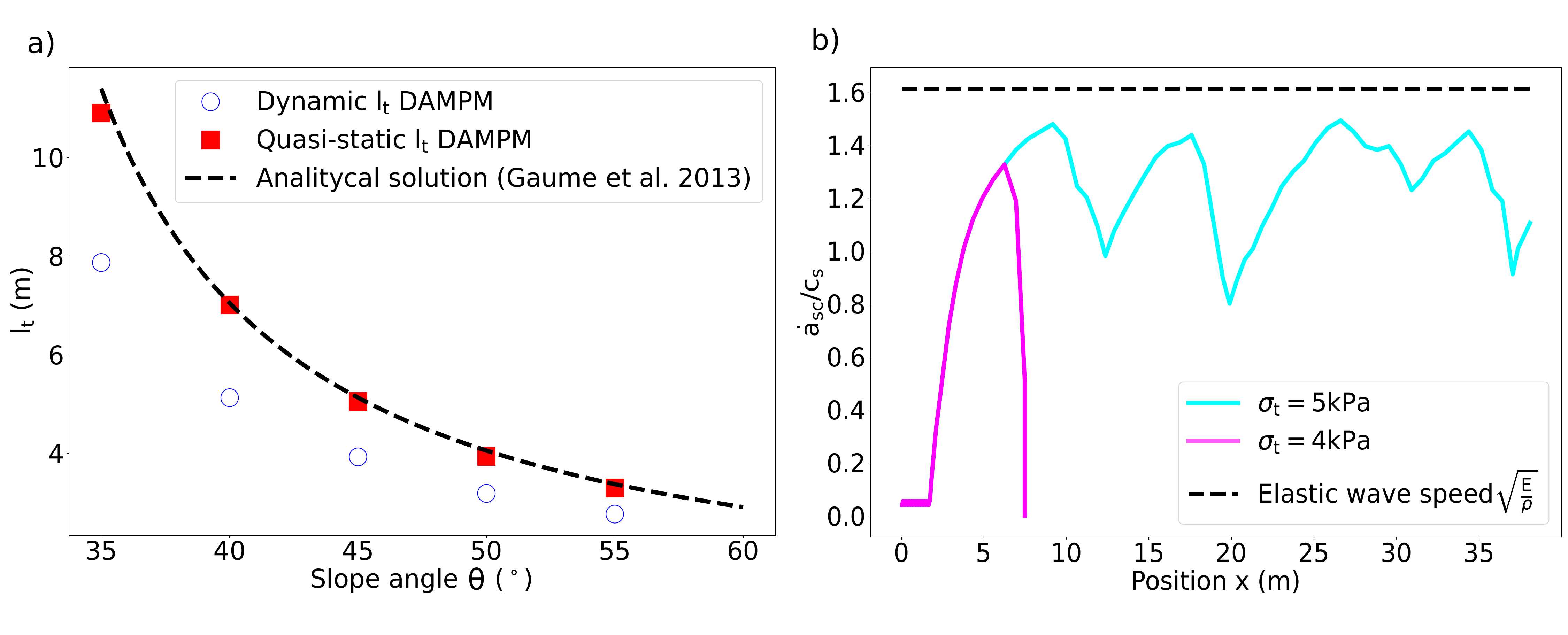}
    \caption{Left: Comparison between the the simulated tensile length in quasi static and dynamic regimes and the analytical tensile length (quasi static). Right: crack velocity profile as a function of the position for two different tensile failure criteria.}
    \label{critical_tensile}
\end{figure}

In the 1D quasi-satic pure shear model, we have an explicit expression of the longitudinal stress in the slab \citep{gaume2013} and therefore have an analytical expression of what must be the tensile fracture length $l_t$. We can now verify the model by comparing the simulated $l_t$ and the analytical one in Fig.~\ref{critical_tensile}a. There are 2 cases in the simulations: either the first tensile fracture occurs during the sawing phase (quasi static regime with $l_t \leq a_{sc}$), or it occurs after the sawing phase (dynamic case with $l_t \geq a_{sc}$). The simulated tensile fracture length in the quasi-static case is very close to the theoretical one, which give us a strong verification for the stress computation within the slab. However, we remark that the simulated tensile fracture length in the dynamic case is lower than the theoretical one. This phenomena occurs one the hand because we neglected the dynamic term in the expression of the theoretical $l_t$ and on the other hand this can be related to the oscillations in the longitudinal stress (see Fig.~\ref{PST_Elastic}).

In addition, if we analyse the velocity profile of the crack tip as a function of the position, we observe mainly two cases depending of the value of $\sigma_t$. Either we have a crack arrest: the propagation of the crack in the WL is stopped when the tensile fracture occurs (magenta curve in Fig.~\ref{critical_tensile} b). If the tensile strength is large enough, the crack continues after the slab fracture. The later slows down the crack velocity, which increases again toward the longitudinal wave speed. 


\subsection{2D crack propagation with an elastic slab}

\begin{figure}[h]
    \centering
    \includegraphics[width=0.85\textwidth]{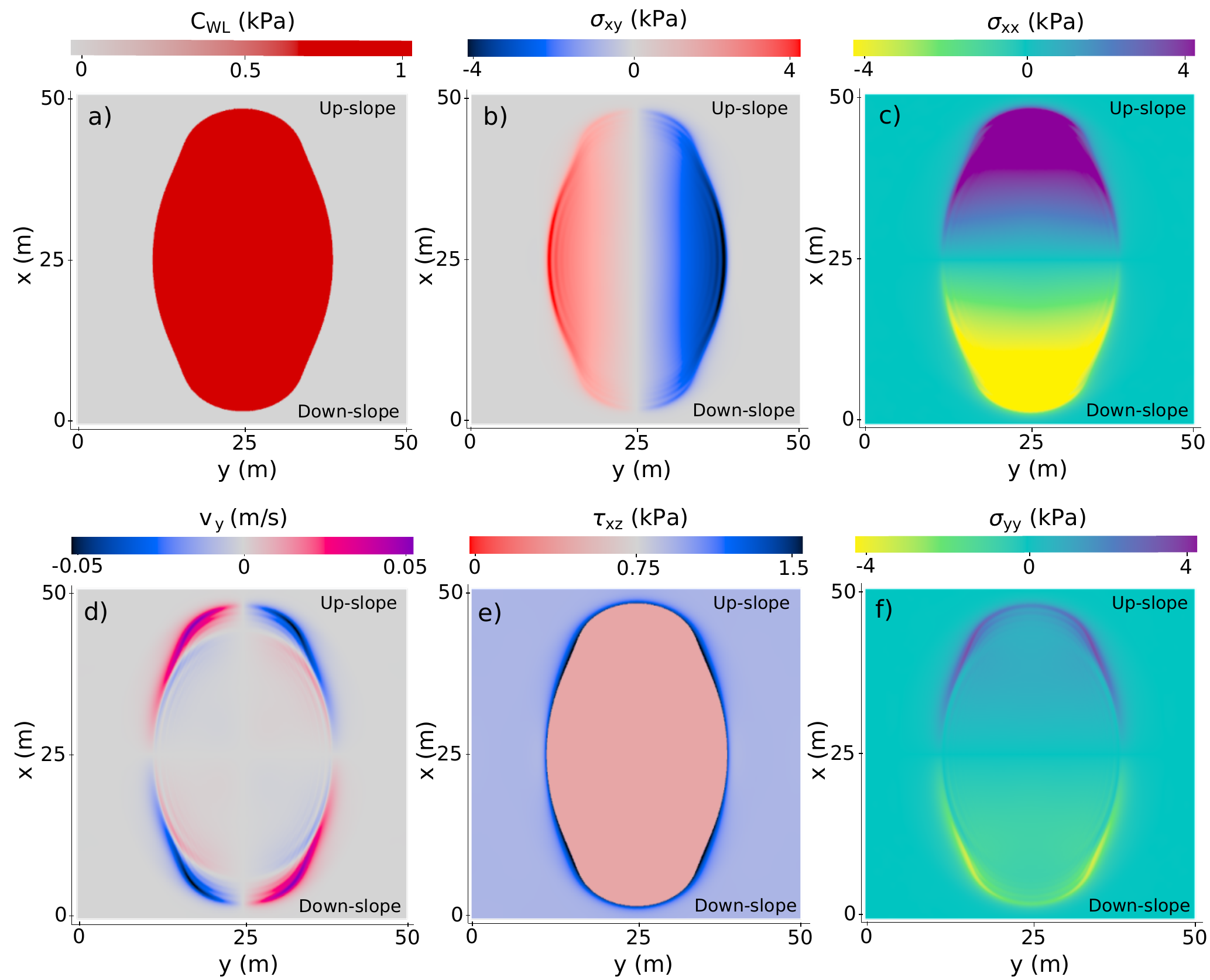}
    \caption{Top view of a different outputs of 2D crack propagation with a purely elastic slab.}
    \label{2d_propagation}
\end{figure}

Previous simulations were performed with one main crack propagation direction (1D). In this section, we perform two dimensional simulations of crack propagation in the weak layer for a purely elastic slab. The output of such simulation at time $t^* = 0.2$s are given in Fig.~\ref{2d_propagation}. The represented quantities are the cohesion (a), the shear stress within the slab (b), the longitudinal stress (down/up slope) (c), the velocity in the slope direction (d), the basal forces (e) and the transversal stress (cross-slope) (f). The parameters used for the simulation are given in case 3 of Table~\ref{PST_elastic_table}. After the initial loading phase, we manually and progressively remove a cohesion circle at the centre of the slab until observing self propagation of the crack in both direction. The crack in the weak layer grows with a shape resembling that of an ellipse. The shape of the crack area appears to converge toward a self similar shape as both cross slope and down-slope and cross-slope crack velocity converges (see also supplementary video 3). From Fig.~\ref{2d_propagation} e), we observe that, as in the PST simulation, the basal force is equal to the frictional stress $\tau_r$ where the cohesion is set to zero. It reaches then $\tau_p$ at the edge of the crack area before converging toward $\tau_g$. We also observe that the crack propagation is mostly driven by the shear within the slab $\sigma_{xy}$ in the cross slope direction and by the longitudinal stress in $\sigma_{xx}$ in the up-slope/down-slope direction. The transversal stress $\sigma_{yy}$ is indeed far less in norm than the other component of the plane stress tensor.

\begin{figure}
    \centering
    \includegraphics[width=0.6\linewidth]{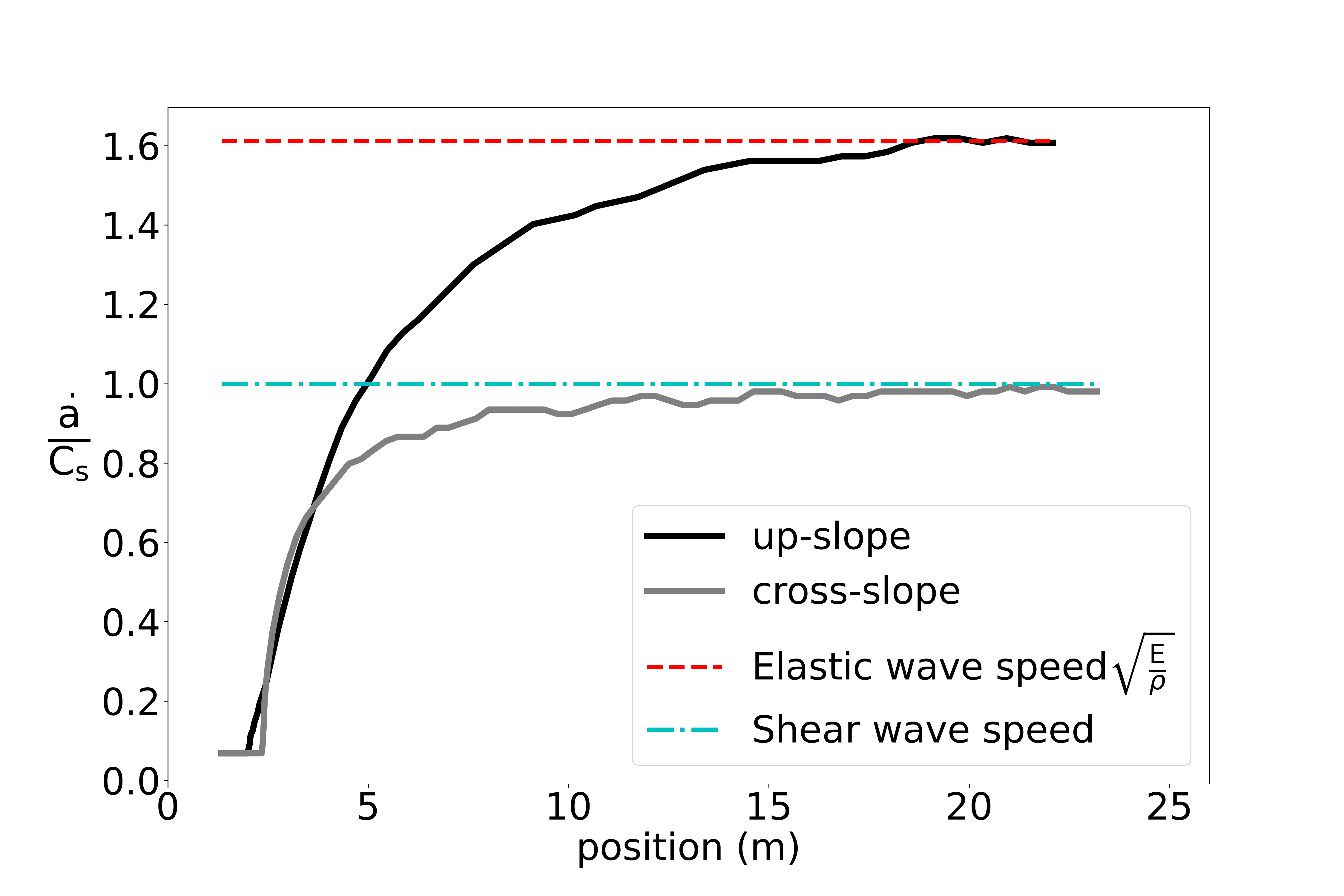}
    \caption{Cross-slope and up-slope crack velocities }
    \label{Velocity_2d}
\end{figure}

If we focus on the crack velocity profile in the $x$ and $y$ direction (Fig.~\ref{Velocity_2d}), we see that the up-slope crack velocity converges to the longitudinal wave speed $c_p$ but the cross-slope crack velocity converges toward $c_s$. This comfort the fact that the up slope crack propagation is driven by the longitudinal wave and the cross-slope propagation is driven by the shear wave. 

\subsection{2D crack propagation with an elastic-brittle slab}

\begin{figure}[H]
    \centering
    \includegraphics[width=1\linewidth]{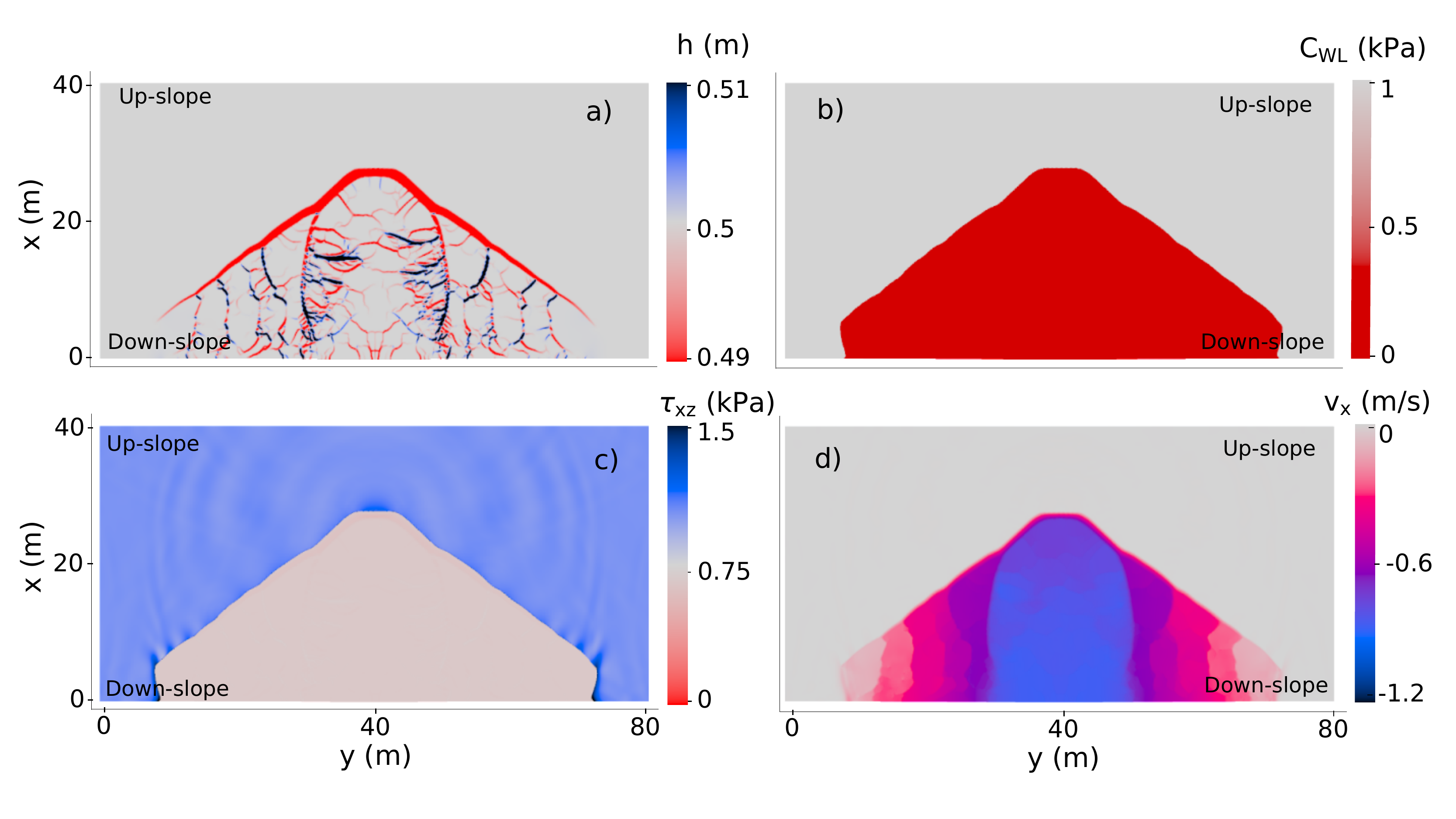}
    \caption{Top view of 3D DAMPM simulation with a plastic model for the slab}
    \label{square_MCC}
\end{figure}

In this section, we relax the hypothesis of an elastic slab and perform simulations in which the slab can break according to the Cohesive Cam Clay model presented in the Methods. This will allow us to evaluate the shape of the avalanche release area. Fig.~\ref{square_MCC} shows four physical outputs of such a simulation during the formation of the release area. The parameters used for this simulation can be found in case 4 of Table~\ref{PST_elastic_table}. We have from left to right the velocity in the slab $v_x$ (a), the weak layer shear stress $\tau_0$ (b), the height (c) and the cohesion (d). The CCC parameters used in this simulation ($\beta = 0.1$, $p_0 = 15$kPa, $M = 1.7$) allowed us to model realistic tensile, shear and compressive strengths (see Methods). In the height output (Fig.~\ref{square_MCC}a), the red color corresponds to tensile failures with a decrease of the height. On the other hand, blue and dark blue colors correspond to compressive failures with an increase in height. We observe that the area below the crown (tensile failure) corresponds to the failed part of the weak layer in which cohesion is zero. In this area, the shear stress in the weak layer is equal to the residual frictional stress and we remark small stress variations induced by local height variations induced by the slab fracture. Above the crown, the stress in the weak layer is almost equal to the shear stress induced by the slab weight $\tau_g$ with small oscillations related to elastic wave propagation (simulation with almost no damping). Similar to the observations made above in the elastic slab case, we have strong shear stress concentrations reaching $\tau_p$ at the weak layer crack tip. Ultimately, the broken slab pieces start to slide with different velocities and a negative velocity gradient from the center of the simulation to its edge.

\section{Discussion}

In this paper, we proposed a depth-averaged Material Point Method (DAMPM) to simulate crack propagation and slab fracture during the release process of dry-snow slab avalanches. This study was motivated by recent findings by \citet{Trottet2022} evidencing a transition from anticrack to supershear fracture after a few meters of propagation in the weak layer. In this supershear regime, slab tension drives the fracture whereas slab bending can be neglected leading to pure shear (mode II) crack propagation in the weak layer. These supershear avalanches propagate with intersonic crack velocities and are believe to generate large avalanche release zones. Hence, for the sake of risk management, in which 100 or 300 year return period avalanches are usually considered, the assumption of weak layer shear failure appears appropriate. The DAMPM model presented in this article is more efficient in terms of computational time compared to the 3D MPM of \citet{Trottet2022}. For instance, the DAMPM PST simulation in case 1 and 2 have around 150'000 particles each and would require more than 1 million particles in 3D. Similarly, the DAMPM simulation shown in case 3 has around 1 million particles and would require 10 times more in 3D. Our approach is thus naturally faster than 3D DEM or FEM models. Compared to limit equilibrium analysis, our approach has the main advantage to be able to capture the dynamic interplay between weak layer failure and slab fracture, thus enabling the evaluation of release zones. 

After verification of the method based on analytic solutions, it was applied to three dimensional terrain with or without enabling slab fractures. It was found that the DAMPM model was able to reproduce the quasi-static prediction of the super critical crack length with a classical shear band propagation approach. In addition, the onset of slab fracture was well reproduced in quasi-static conditions. Finally, it was verified that the crack propagation speed in the up/down slope direction was reaching the longitudinal wave speed $c_P$. On the other hand, the cross slope crack speed converges to $c_S$. This leads to an elliptical propagation shape. With their depth-averaged finite difference model, \citep{Zhang2022} reported different types of shapes. By restricting the motion in the lateral direction, they also report an elliptical shape. However, without this restriction, they report a `peanut' propagation shape. This particular shape is due to the non-symmetry of the lateral slab velocity. Surprisingly, while this non-symmetry is also present in our simulations, a `peanut' shape was initially not observed. It is only when the scale of the simulation was increased or if the characteristic length $\Lambda$ was decreased (e.g. through a decrease of the slab elastic modulus) that this peculiar shape was recovered. As a consequence, the shape of the ellipse is not self-similar in time. This is induced by the mismatch of the longitudinal and lateral crack propagation speeds.

\begin{figure}
     \centering
         \includegraphics[width=\textwidth]{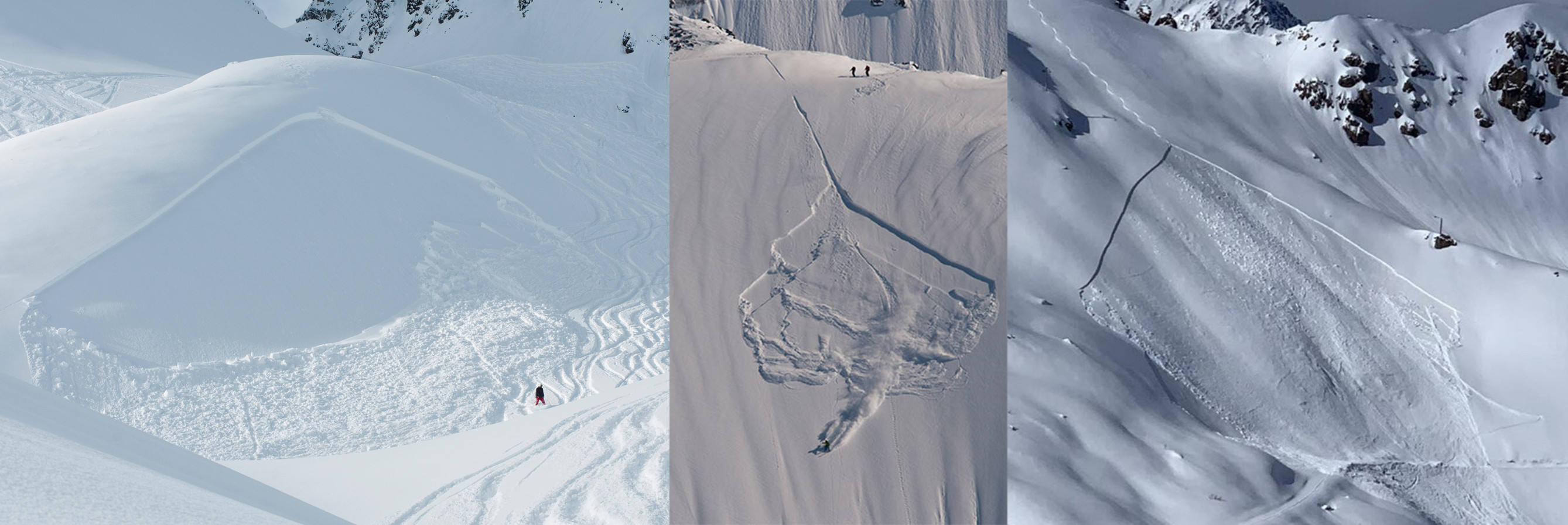}
        \caption{Avalanche release photographs. Credit from left to right: Rémi Petit, montagne-magazine.com and Isabelle Sabater.}
        \label{Real_ava}
\end{figure}

It is important to note that, in principle, based on theoretical considerations \citep{Broberg1989}, the Rayleigh wave speed $c_R$ should not be exceeded in Mode II. In our study, slab bending and shear within the depth are not taken into account due to the plane stress assumption made for depth-averaging. Hence, in our model, there is no reason why the propagation speed should be limited by $c_R$. This is why in our case, we have a smooth increase of the speed towards the longitudinal wave speed $c_P$. In 3D simulations, a daughter supershear crack nucleates ahead of the main fracture with a speed which continuously increases towards $c_P$, while crossing the forbidden region between $c_R$ and $c_S$ \citep{Gaume2019,bergfeld2022,Bizzarri2012}. We obviously cannot reproduce the discontinuity reported in 3D simulations, but the temporal evolution of the supershear crack speed is in line with the latter 3D simulation. Furthermore, in mode III (cross-slope), we obtain a convergence to $c_S$ in line with previous theoretical work \citep{Broberg1996}.

\begin{figure}[h]
     \centering
         \includegraphics[width=\textwidth]{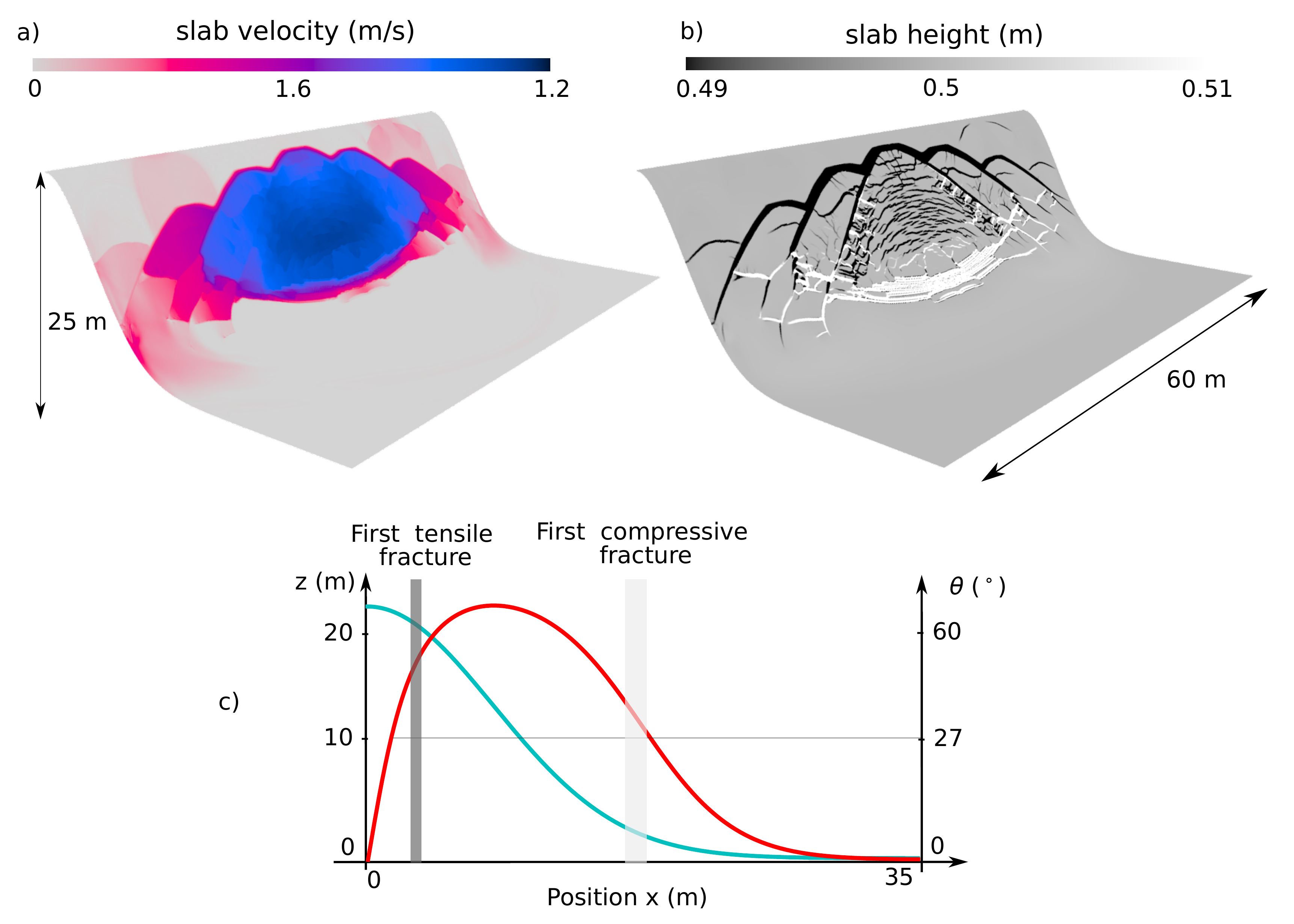}
        \caption{Simulation outputs with variation of the topography. a) Tangential velocity distribution in the slab. b) height distribution in the slab. c) altitude profile (blue) and slope angle profile (red) at the middle of the width with the first tensile/compressive failure marked.}
        \label{Quasi_1d}
\end{figure}

Allowing for slab mixed-mode fracture in DAMPM allowed us to obtain interesting `hand fan' shape for the release area. In details, the half angle at the top of the fan is around 50 degrees for the simulation in Fig.~\ref{square_MCC}. While we cannot claim validation, this shape is clearly in line with field observations of artificially triggered avalanches as seen in Fig. \ref{Real_ava} in which the half fan angle appears to be between 40 and 50 degrees. In our model, the detailed shape is influenced by the choice of plastic parameters, but the overall shape remains similar. We also performed a larger simulation on a wider slope including a convex top ridge and concave transition to flat terrain (Fig. \ref{Quasi_1d}). Interestingly, this simulation shows a race between crack propagation in the weak layer and slab fractures, leading to lateral en-echelon failures, a process often reported in the field \citep{bergfeld2022}. The overall shape differs from the pictures in Fig. \ref{Real_ava}, but is in line with observation of larger avalanches in which the crown fracture is very often irregular and impacted by topography. In this simulation, it is interesting to report that the tensile crown failure occurs for a slope angle around 35 degrees and that the stauchwall failure is occurring for an angle very close to the friction angle of the weak layer.

Our model has strong similarities with the cellular automata model developed by \citep{Fyffe2004, Fyffe2007}. In their study, they were able to simulate both weak layer shear failure and slab fracture, based on finite differences. The weak layer model is almost identical but their slab constitutive law does not consider mixed-mode failure. Together with a different numerical scheme used, the different slab model is very likely the reason why the release shapes they obtain differ from ours. In particular, they report a release zone which is larger in the downslope direction than in the cross slope direction. This contrasts with our results and with field observation of avalanche release zones made by \cite{McClung2009}. The latter author reports a ratio between cross-slope and down/up slope dimensions of the release area to be around 2.1 for unconfined conditions. The shape of our release area also differs from that of \cite{Zhang2022} with a mismatch very likely due to a different slab constitutive model (von Mises versus CCC).

Concerning numerical limitations, our DAMPM implementation is based on a FLIP/PIC transfer scheme where particles quantities are transferred to grid nodes using cubic splines with a support of $2 \Delta x$.  With this particular scheme, a balance must be found between accuracy (pure FLIP) and stability (pure PIC). Here we chose to promote accuracy with a FLIP/PIC ratio of 0.9 and thus limited damping. The numerical oscillations increase with the FLIP/PIC ratio and by increasing the ratio $\Delta t/ \Delta x$. Such oscillations in space can be seen in the longitudinal stress profile in Fig.~\ref{PST_tensile}. More advanced interpolation schemes, such as APIC~\citep{JIANG}, could improve accuracy while preserving stability. 

Finally, most of the presented simulation outputs have been found independent of the mesh resolution (critical crack length, distance for slab fracture, crack propagation speed, avalanche shape) provided the grid size $\Delta x$ is small than the characteristic elastic length $\Lambda$. In addition, as expected and reported from 3D simulations, the arrest of crack propagation in the weak layer during slab fracture (Fig.~\ref{critical_tensile}b) is mesh-dependent. This is related to the plastic strain-softening behaviour of the slab. In the future, a regularised model could lead to some improvements in this direction \citep{Mahajan2010, SULSKY20111674}.


\section{Conclusion and outlook}

Motivated by recent work revealing a transition from anticrack to supershear crack propagation, we developed a depth-averaged Material Point Method with a weak layer shear failure model to evaluate large-scale crack propagation mechanisms and avalanche release areas. We checked the validity of our model by simulating simple test cases for which analytic solutions exist. We then applied the model on slope-scale simulations and obtained avalanche release shapes qualitatively in good agreement with field observations. In the future, more complex topographies and snowpack spatial variability, for instance based on remote sensing data, should be considered to bring the proposed model at an operational state for automatic evaluation of avalanche release zones in alpine regions. This achievement could contribute to define avalanche size indices for avalanche forecasting and also serve as input of avalanche flow models used for risk assessment. Finally, it would be straightforward to modify the constitutive model and interface friction law assumed here to be able to simulate different processes, for instance the dynamics of avalanches or other types of gravitational mass movements, snow creep and its impact on structures, glacier flow and calving.


\appendix

\section{Depth integration on complex topography}
\label{complex_terr}

In the previous integration, we supposed that the terrain was flat. In order to simulate the material on a complex terrain we have to change the previous equation by adding in the depth-averaged momentum conservation some terms depending on the surface gradients. However, for an arbitrary topography, it is more relevant to pass from cartesian coordinates to curvilinear coordinates, adapted to the topography. This will allow us to integrate the equations according to the surface normal. For the change of coordinates and the integration which follows, we shall use the same notation as the one introduced in 
\citet{Vila2008} and \citet{Bouchut2004}.
Thus the notation $\phi_n$ denote the vertical component of a vector $\vec{\phi} = \begin{pmatrix} \bm \phi \\ \phi_n \end{pmatrix}$, where $\bm \phi$ is the tangential vector.

\begin{figure}[h]
    \centering
    \includegraphics[width=10cm]{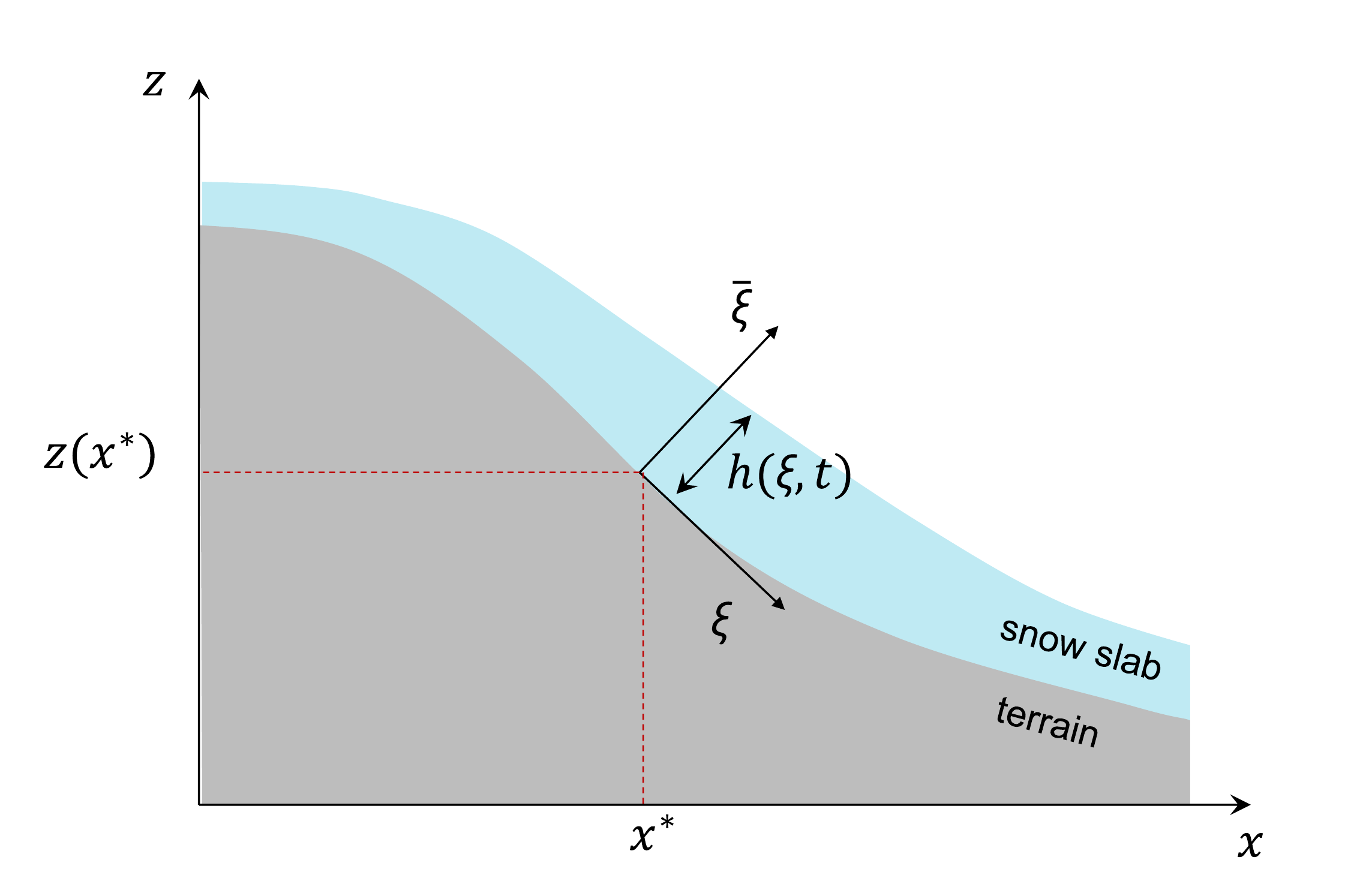}
    \caption{illustration of the new set of variable $(\xi, \xi_n)$ following the slope topography}
    \label{complex}
\end{figure}

We denote $\bm x \in \mathbb{R}^N$ the horizontal cartesian coordinates ($N = 1$ or $N = 2$ according of the modeling) and $\bm \xi \in \mathbb{R}^N$ the tangential curvilinear coordinates which can be seen as of function of the variable $\bm x$. The $N \times N$ matrix $\dif{\bm \xi}{\bm x}$ represent the jacobian matrix of the function $\bm x$. 

If the bottom surface $\mathcal{S}$ is the graph of a function $z : \mathbb{R}^N \longmapsto \mathbb{R}$ , i.e. $\mathcal{S} = \Big\{ \big( \bm x, z(\bm x) \big), \bm x \in \mathbb{R}^N \Big\}$, we can easily define the normal $\bm n $ to $\mathcal{S}$ as 
$$
\bm n = \left( \begin{array}{c}
     - \bm s  \\
    c 
\end{array} \right), \quad \bm s  = (1 + \| \nabla_x z \|^2)^{\frac{-1}{2}} \nabla_x z, \quad c = (1 + \| \nabla_x z \|^2)^{\frac{-1}{2}}
$$

In the previous equation, $\nabla_{x}$ represents the gradient with respect to the variable $\bm x$. note that since $\bm n$ is unitary, we have the following relation $\| \bm n \|^2 = \| \bm s \|^2 + c^2 = 1$. 
This leads to the following equation 
\begin{equation}
    \dif{x}{c} = - \frac{1}{c}\bm s^t \dif{x}{\bm s} , \quad \dif{x}{\bm s } = c(1 - \bm s^t \bm s ) \partial_{xx}^2 z , \quad \partial_{xx}^2 z = \frac{c^2Id + \bm s \bm s^t}{c^3} \dif{x}{\bm s}
    \label{fgeom}
\end{equation}

We then define a new system of coordinate $\vec{\xi} = (\bm \xi, \xi_n)$ where $\xi_n$ is the coordinate along the normal to the bed $\mathcal{S}$. If $\Omega_t$ is the fluid domain at time $t$, we have
$$
\vec{X} (\vec{ \xi}) \in \Omega_t \iff \vec{X} (\vec{ \xi}) = \left( \begin{array}{c}
     \bm x(\bm \xi) \\
     z(\bm x(\bm \xi)) 
\end{array}\right) + \xi_n \; \bm n (\bm x(\bm \xi)), \quad 0 \leq \xi_n \leq h(\bm \xi, t)
$$

We can now define new velocity components $\vec{V}$ with respect of the new coordinates, if we denote by $\vec{U}$ the velocity field in the Cartesian base, we have
\begin{equation}
    \vec{U} = \big( \partial_{\vec{\xi}} \vec{X} \big) \vec{V} 
\end{equation}
with 
\begin{equation}
    \partial_{\vec{\xi}}\vec{X} = \begin{pmatrix}
        \partial_{\xi} X & - \bm s \\
        \frac{1}{c} \bm s ^t (\partial_{\xi} X) & c \\
    \end{pmatrix} \in \mathcal{M}_{n+1}(\mathbb{R})
    \label{change_vec}
\end{equation}
and 
\begin{equation}
    \partial_{\xi} X = (Id - \xi_n \dif{x}{\bm s}) \dif{\xi}{\bm x} \in \mathcal{M}_{n}(\mathbb{R})
\end{equation}

We denote $A = \big(\partial_{\vec{\xi}}\vec{X}\big)^{-1} \in \mathcal{M}_{n+1}(\mathbb{R})$.  Inserting Eqs.~\eqref{change_vec} and \ref{fgeom}, we have 
\begin{equation}
    A = \begin{pmatrix}
        \big(\partial_{\xi} X\big)^{-1} & 0 \\
        0 & 1 \\
    \end{pmatrix} \begin{pmatrix}
        Id - \bm s \bm s^t & c \bm s \\
        - \bm s ^t (\partial_{\xi} X) & c \\
    \end{pmatrix}
\end{equation}
We can then compute the matrix $\Tilde{M} = A A^t \in \mathcal{M}_{n+1}(\mathbb{R})$, which can be rewritten as
\begin{equation}
    \Tilde{M} = \begin{pmatrix}
        M & 0 \\
        0 & 1 \\
    \end{pmatrix} , \quad M = (\dif{\xi}{X})^{-1}(Id - \bm s \bm s^t)(\dif{\xi}{X})^{-t} 
\end{equation}
and we now compute the jacobian $J$ of the coordinates transformation:
\begin{align}
    J = \det (\dif{\vec{\xi}}{\vec{X}}) = \frac{1}{c} \det (\dif{\xi}{X}) = \det(M)^{-\frac{1}{2}}.
\end{align}

The change of variable will induce a change of the differentiable operator used in the governing equations. This result is the Lemma 1 of \citet{Vila2008} and the proof can be found in \citet{Bouchut2004}. For any field $\vec{Z}$, the differential operator transform to 
\begin{equation}
    J \nabla_{\vec{X}} \cdot \vec{Z} = \nabla_{\vec{\xi}} \cdot (J A\vec{Z}), \quad \vec{U} \cdot \nabla_{\vec{X}} = \vec{V} \cdot \nabla_{\vec{\xi}}, \quad \nabla_{\vec{X}} = A^t \nabla_{\vec{\xi}}
    \label{transform_vec}
\end{equation}
and for any symmetric tensor $\sigma$,
\begin{equation}
    J A^{-t} \nabla_{\vec{X}} \cdot \sigma = \nabla_{\vec{\xi}}\big( J \mathcal{P} A A^t\big) + \frac{J}{2} \mathcal{P} : \nabla_{\vec{\xi}}\Tilde{M}, \quad \text{with} \quad \mathcal{P} = A^{-t} \sigma A^{-1}
    \label{transform_tensor}
\end{equation}
We notice that the tensor $\mathcal{P}$ is the stress tensor oriented in the curvilinear coordinates. 

\paragraph*{Boundary conditions}

We have the same kinematic condition as with the cartesian coordinates, $i. e.$
\begin{equation}
    \dif{t}{h} + V_h \cdot \nabla_{\xi} h = \overline{V}_h
\end{equation}
with the following notation for a field $\phi$, $\phi_h(\xi) = \phi(\xi, h(\xi, t))$. \\
The stress free condition yields that for $\overline{\xi} = h(\xi, t)$, no forces are apply on the slab, this means 
\begin{equation}
    \sigma_h = 0
\end{equation}

\paragraph*{Governing equation in new coordinates}

In the incompressible case, the mass conservation is reduced to $\nabla_{\vec{X}} \cdot \vec{U} = 0$. Multiplying this equation by $J$ and use equation \ref{transform_vec}, we get
\begin{equation}
    \nabla_{\vec{\xi}} \cdot (J\vec{V}) = \nabla_{\xi} \cdot (J V) + \dif{\overline{\xi}}{J \overline{V}} = 0
    \label{mass_complex}
\end{equation}

The equation translating the momentum conservation is, in cartesian coordinates
\begin{equation}
    \dif{t}{\vec{U}} + \vec{U} \cdot \nabla_{\vec{X}} \vec{U} = -\vec{g} + \frac{1}{\rho} \nabla_{\vec{X}} \cdot \sigma.
    \label{momentum_villa}
\end{equation}
If we multiply Eq.~\eqref{momentum_villa} by matrix $A$ and use the curviliear operators, one finds 
\begin{equation}
    \dif{t}{\vec{V}} + \vec{V} \cdot \nabla_{\vec{\xi}} \vec{V} = -A \vec{g} + \vec{\Gamma}(\vec{V}) +  \frac{1}{\rho J} \Tilde{M} \big(\nabla_{\vec{\xi}} \cdot (J\mathcal{P} \Tilde{M}) + \frac{J}{2} \mathcal{P} : \nabla_{\vec{\xi}} \Tilde{M} \big)
    \label{prime}
\end{equation}
The term $\vec{\Gamma}(\vec{V})$ is a correction induced by the change of coordinate in the convective term, we have
\begin{equation}
    \vec{\Gamma}(\vec{V}) = A \vec{V} \cdot \nabla_{\vec{\xi}}(A^{-1} \vec{V}) - \vec{V} \cdot \nabla_{\vec{\xi}} \vec{V}
    \label{qdm_complex}
\end{equation} 
This term will be neglected in the next section. In order to simplify the notation and highlight the different component of the tensor stress $\mathcal{P}$, we decompose
\begin{equation}
    \mathcal{P} = \begin{pmatrix}
        P & Z \\
        Z^t & \sigma_{\overline{\xi}} \\
    \end{pmatrix} 
\end{equation}
$P_{ij}$ corresponds to $\xi_i \xi_j$ direction of the stress tensor and $Z$ corresponds to the shear in the $\xi_i \overline{\xi}$ direction.

If we projected Eq.~\eqref{qdm_complex} in the $(\xi_1, ... , \xi_n)$ direction we have 
\begin{equation}
    \dif{t}{V} + \vec{V} \cdot \nabla_{\xi} V = -g c \big( \dif{\xi}{X}\big)^{-1} \bm s + \Gamma(\vec{V}) + \frac{1}{\rho J} M \big( \nabla_{\xi} \cdot (J P M) + \dif{\overline{\xi}}{J Z} + \frac{J}{2} P : \nabla_{\xi} M \big)
    \label{qdm_complex_1}
\end{equation}

The latter equation is almost the same equation as the momentum conservation equation in curvilinear coordinates found in \citep{Vila2008}. However, in order to model an elasto-plastic behaviour of the slab, no further hypothesis will be assumed on the stress tensor $\mathcal{P}$, which will be namely the one use in the elastic constitutive law.

\subsubsection{Shallow water and small velocity hypothesis}

We can express the shallow water hypothesis with the new set of coordinates \citep{Bouchut2004}, which are namely
\begin{itemize}
    \item the height of the material is small comparing to $\varepsilon$, $h = \mathcal{O}(\varepsilon)$
    \item the curvature is small comparing to $\varepsilon$, $(\partial^2_{xx} z) = \mathcal{O}(\varepsilon)$
    \item the velocity field does not depend on the depth of the material, $\vec{V}(t, \vec{\xi}) = V(t, \bm \xi) + \mathcal{O}(\varepsilon)$
\end{itemize}

With this hypothesis, we have 
$$
\dif{\bm \xi}{\bm X} = \dif{\bm \xi }{\bm x} + \mathcal{O}(\varepsilon^2) 
$$
which implies that 
\begin{equation}
    J = \overline{J} + \mathcal{O}(\varepsilon^2), \quad M = \overline{M} + \mathcal{O}(\varepsilon^2) 
\end{equation}
with
\begin{equation}
    \overline{M} = (\dif{\xi}{\bm x})^{-1}(Id - \bm s \bm s^t)(\dif{\xi}{\bm x})^{-t}, \quad \overline{J} = \det (\overline{M})^{-\frac{1}{2}}.
    \label{simplification}
\end{equation}
Eq.~\eqref{qdm_complex_1} can thus be rewritten as 
\begin{equation}
    \dif{t}{V} + \vec{V} \cdot \nabla_{\xi} V = -g c \big( \dif{\xi}{\bm x}\big)^{-1} \bm s + \Gamma(\vec{V}) + \frac{1}{\rho \overline{J}} \overline{M} \big( \nabla_{\xi} \cdot (J P \overline{M}) + \dif{\overline{\xi}}{\overline{J} Z} + \frac{\overline{J}}{2} P : \nabla_{\xi} \overline{M} \big).
    \label{qdm_simp}
\end{equation}

\paragraph{Small velocity} In this work, we want to simulate initiation of snow slab avalanches. In this process, we have a fast propagation of stress within the slab but the particles in slab have a very small velocity. The term $\Gamma(\vec{V})$ varies as the square of the velocity norm and thus will be neglected in this framework.  

\subsubsection{Depth-averaged equation for the quasi-1D case}

The integration of Eq.~\eqref{qdm_complex} is strongly inspired from \citep{Vila2008} and \citep{Bouchut2004}. In this section, we will suppose the the terrain varies only in one dimension (we suppose $x$). The function $z$ representing the height of the bed does not depend of the variable $y$, i.e.,
$$
\forall x, \; y \in \mathbb{R}, \der{z}{y}(x, y) = 0.
$$
We denote then $\theta = \arctan \Big(  \der{z}{x} \Big)$ such that $\dif{\xi}{\bm x}$ can be rewritten as 

\begin{equation}
    \dif{\xi}{\bm x} = \begin{pmatrix}
        \cos \theta & 0 \\
        0 & 1 \\
    \end{pmatrix}, \quad \text{and} \quad \bm s = \begin{pmatrix} 
        \sin \theta \\
        0
    \end{pmatrix}
\end{equation}
Inserting this in Eq.~\eqref{simplification}, we have 
\begin{equation}
    \olM = Id \quad \olJ = \det(\olM) = 1 
\end{equation}
Eq.~\eqref{qdm_simp} becomes then
\begin{equation*}
    \dif{t}{V} + \vec{V} \cdot \nabla_{\xi} V = -g c \big( \dif{\xi}{\bm x}\big)^{-1} \bm s + \Gamma(\vec{V}) + \frac{1}{\rho}\big( \nabla_{\xi} \cdot P + \dif{\overline{\xi}}{Z} \big).
\end{equation*}
Finally, by neglecting the term $\Gamma(\vec{V})$, the system becomes
\begin{equation}
    \dif{t}{V} + \vec{V} \cdot \nabla_{\xi} V = -g c \big( \dif{\xi}{\bm x}\big)^{-1} \bm s + \frac{1}{\rho}\big( \nabla_{\xi} \cdot P + \dif{\overline{\xi}}{Z} \big).
    \label{qdm_proj_simp}
\end{equation}

Eq.~\eqref{qdm_proj_simp} has the exact same form as Eqs.~\eqref{non_conservative_x_inc} and \eqref{non_conservative_y_inc}. Therefore, the integrated equation will have the exact same form, namely 
\begin{align}
    \dif{t}{h} + \nabla_{\xi} (h \olv) &= 0 \\
    \dif{t}{h \olv} + \nabla_{\xi} \big(h \olv \times \olv \big)  &= - h g c \big( \dif{\xi}{\bm x}\big)^{-1} \bm s + \frac{1}{\rho} \nabla_{\xi} \cdot \big( h \ols \big) + \frac{1}{\rho} Z_0.
    \label{conservative_qdm_Simp}
\end{align}
with 
\begin{equation*}
    \olv = \int_0^{h(\xi, t)} V  d\overline{\xi},   \quad \ols =  \int_0^{h(\xi, t)} P  d\overline{\xi}, \quad Z_0 = Z(\overline{\xi} = 0).
\end{equation*}
Note that we have an explicit form for the product $c \big( \dif{\xi}{\bm x}\big)^{-1} \bm s$
\begin{equation*}
    c \big( \dif{\xi}{\bm x}\big)^{-1} \bm s = \bm s = \begin{pmatrix}
        \sin \theta \\
        0
    \end{pmatrix}
\end{equation*}

\acknowledgments
This work was supported by the Swiss National Science Foundation (grant number PCEFP2\_181227).


\begin{thebibliography}{}

\bibitem [\protect \citeauthoryear {%
Abe%
\ \BBA {} Konagai%
}{%
Abe%
\ \BBA {} Konagai%
}{%
{\protect \APACyear {2016}}%
}]{%
Abe16}
\APACinsertmetastar {%
Abe16}%
\begin{APACrefauthors}%
Abe, K.%
\BCBT {}\ \BBA {} Konagai, K.%
\end{APACrefauthors}%
\unskip\
\newblock
\APACrefYearMonthDay{2016}{}{}.
\newblock
{\BBOQ}\APACrefatitle {Numerical simulation for runout process of debris flow
  using depth-averaged material point method} {Numerical simulation for runout
  process of debris flow using depth-averaged material point method}.{\BBCQ}
\newblock
\APACjournalVolNumPages{Soils and Foundations}{56}{}{869--888}.
\newblock
\begin{APACrefDOI} \doi{10.1016/j.sandf.2016.08.011} \end{APACrefDOI}
\PrintBackRefs{\CurrentBib}

\bibitem [\protect \citeauthoryear {%
Ancey%
}{%
Ancey%
}{%
{\protect \APACyear {2006}}%
}]{%
Ancey2006}
\APACinsertmetastar {%
Ancey2006}%
\begin{APACrefauthors}%
Ancey, C.%
\end{APACrefauthors}%
\unskip\
\newblock
\APACrefYear{2006}.
\newblock
\APACrefbtitle {Dynamique des avalanches} {Dynamique des avalanches}.
\newblock
\APACaddressPublisher{}{PPUR presses polytechniques}.
\PrintBackRefs{\CurrentBib}

\bibitem [\protect \citeauthoryear {%
Bergfeld%
\ \protect \BOthers {.}}{%
Bergfeld%
\ \protect \BOthers {.}}{%
{\protect \APACyear {2022}}%
}]{%
bergfeld2022}
\APACinsertmetastar {%
bergfeld2022}%
\begin{APACrefauthors}%
Bergfeld, B.%
, van Herwijnen, A.%
, Bobillier, G.%
, Larose, E.%
, Moreau, L.%
, Trottet, B.%
\BDBL {}et al.%
\end{APACrefauthors}%
\unskip\
\newblock
\APACrefYearMonthDay{2022}{}{}.
\newblock
{\BBOQ}\APACrefatitle {Crack propagation speeds in weak snowpack layers} {Crack
  propagation speeds in weak snowpack layers}.{\BBCQ}
\newblock
\APACjournalVolNumPages{Journal of Glaciology}{68}{269}{557–570}.
\newblock
\begin{APACrefDOI} \doi{10.1017/jog.2021.118} \end{APACrefDOI}
\PrintBackRefs{\CurrentBib}

\bibitem [\protect \citeauthoryear {%
Bizzarri%
\ \BBA {} Das%
}{%
Bizzarri%
\ \BBA {} Das%
}{%
{\protect \APACyear {2012}}%
}]{%
Bizzarri2012}
\APACinsertmetastar {%
Bizzarri2012}%
\begin{APACrefauthors}%
Bizzarri, A.%
\BCBT {}\ \BBA {} Das, S.%
\end{APACrefauthors}%
\unskip\
\newblock
\APACrefYearMonthDay{2012}{}{}.
\newblock
{\BBOQ}\APACrefatitle {Mechanics of 3-D shear cracks between Rayleigh and shear
  wave rupture speeds} {Mechanics of 3-d shear cracks between rayleigh and
  shear wave rupture speeds}.{\BBCQ}
\newblock
\APACjournalVolNumPages{Earth and Planetary Science
  Letters}{357-358}{}{397-404}.
\newblock
\begin{APACrefDOI} \doi{10.1016/j.epsl.2012.09.053} \end{APACrefDOI}
\PrintBackRefs{\CurrentBib}

\bibitem [\protect \citeauthoryear {%
Blatny%
, Berclaz%
, Guillard%
, Einav%
\BCBL {}\ \BBA {} Gaume%
}{%
Blatny%
\ \protect \BOthers {.}}{%
{\protect \APACyear {2022}}%
}]{%
blatny2022}
\APACinsertmetastar {%
blatny2022}%
\begin{APACrefauthors}%
Blatny, L.%
, Berclaz, P.%
, Guillard, F.%
, Einav, I.%
\BCBL {}\ \BBA {} Gaume, J.%
\end{APACrefauthors}%
\unskip\
\newblock
\APACrefYearMonthDay{2022}{}{}.
\newblock
{\BBOQ}\APACrefatitle {Microstructural Origin of Propagating Compaction
  Patterns in Porous Media} {Microstructural origin of propagating compaction
  patterns in porous media}.{\BBCQ}
\newblock
\APACjournalVolNumPages{Phys. Rev. Lett.}{128}{}{228002}.
\newblock
\begin{APACrefDOI} \doi{10.1103/PhysRevLett.128.228002} \end{APACrefDOI}
\PrintBackRefs{\CurrentBib}

\bibitem [\protect \citeauthoryear {%
Blatny%
, Löwe%
, Wang%
\BCBL {}\ \BBA {} Gaume%
}{%
Blatny%
\ \protect \BOthers {.}}{%
{\protect \APACyear {2021}}%
}]{%
blatny2021}
\APACinsertmetastar {%
blatny2021}%
\begin{APACrefauthors}%
Blatny, L.%
, Löwe, H.%
, Wang, S.%
\BCBL {}\ \BBA {} Gaume, J.%
\end{APACrefauthors}%
\unskip\
\newblock
\APACrefYearMonthDay{2021}{}{}.
\newblock
{\BBOQ}\APACrefatitle {Computational micromechanics of porous brittle solids}
  {Computational micromechanics of porous brittle solids}.{\BBCQ}
\newblock
\APACjournalVolNumPages{Computers and Geotechnics}{140}{}{104284}.
\newblock
\begin{APACrefDOI} \doi{10.1016/j.compgeo.2021.104284} \end{APACrefDOI}
\PrintBackRefs{\CurrentBib}

\bibitem [\protect \citeauthoryear {%
Bobillier%
\ \protect \BOthers {.}}{%
Bobillier%
\ \protect \BOthers {.}}{%
{\protect \APACyear {2021}}%
}]{%
Bobillier2021}
\APACinsertmetastar {%
Bobillier2021}%
\begin{APACrefauthors}%
Bobillier, G.%
, Bergfeld, B.%
, Dual, J.%
, Gaume, J.%
, van Herwijnen, A.%
\BCBL {}\ \BBA {} Schweizer, J.%
\end{APACrefauthors}%
\unskip\
\newblock
\APACrefYearMonthDay{2021}{}{}.
\newblock
{\BBOQ}\APACrefatitle {Micro-mechanical insights into the dynamics of crack
  propagation in snow fracture experiments} {Micro-mechanical insights into the
  dynamics of crack propagation in snow fracture experiments}.{\BBCQ}
\newblock
\APACjournalVolNumPages{Scientific Reports}{11}{}{2045-2322}.
\newblock
\begin{APACrefDOI} \doi{10.1038/s41598-021-90910-3} \end{APACrefDOI}
\PrintBackRefs{\CurrentBib}

\bibitem [\protect \citeauthoryear {%
Bouchut%
\ \BBA {} Westdickenberg%
}{%
Bouchut%
\ \BBA {} Westdickenberg%
}{%
{\protect \APACyear {2004}}%
}]{%
Bouchut2004}
\APACinsertmetastar {%
Bouchut2004}%
\begin{APACrefauthors}%
Bouchut, F.%
\BCBT {}\ \BBA {} Westdickenberg, M.%
\end{APACrefauthors}%
\unskip\
\newblock
\APACrefYearMonthDay{2004}{}{}.
\newblock
{\BBOQ}\APACrefatitle {Gravity driven shallow water model for arbitrary
  topography} {Gravity driven shallow water model for arbitrary
  topography}.{\BBCQ}
\newblock
\APACjournalVolNumPages{Communications in Mathematical Sciences}{3}{}{359-389}.
\newblock
\begin{APACrefDOI} \doi{cms/1109868726} \end{APACrefDOI}
\PrintBackRefs{\CurrentBib}

\bibitem [\protect \citeauthoryear {%
Boutounet%
, Chupin%
, Noble%
\BCBL {}\ \BBA {} Vila%
}{%
Boutounet%
\ \protect \BOthers {.}}{%
{\protect \APACyear {2008}}%
}]{%
Vila2008}
\APACinsertmetastar {%
Vila2008}%
\begin{APACrefauthors}%
Boutounet, M.%
, Chupin, L.%
, Noble, P.%
\BCBL {}\ \BBA {} Vila, J\BHBI P.%
\end{APACrefauthors}%
\unskip\
\newblock
\APACrefYearMonthDay{2008}{}{}.
\newblock
{\BBOQ}\APACrefatitle {Shallow water viscous flows for arbitrary topography}
  {Shallow water viscous flows for arbitrary topography}.{\BBCQ}
\newblock
\APACjournalVolNumPages{Communications in Mathematical Science}{}{}{}.
\newblock
\begin{APACrefDOI} \doi{10.1016/j.ijsolstr.2017.12.033} \end{APACrefDOI}
\PrintBackRefs{\CurrentBib}

\bibitem [\protect \citeauthoryear {%
Broberg%
}{%
Broberg%
}{%
{\protect \APACyear {1989}}%
{\protect \APACexlab {{\protect \BCnt {1}}}}}]{%
Broberg1996}
\APACinsertmetastar {%
Broberg1996}%
\begin{APACrefauthors}%
Broberg, K\BPBI B.%
\end{APACrefauthors}%
\unskip\
\newblock
\APACrefYearMonthDay{1989{\protect \BCnt {1}}}{}{}.
\newblock
{\BBOQ}\APACrefatitle {How fast can a crack go?} {How fast can a crack
  go?}{\BBCQ}
\newblock
\APACjournalVolNumPages{Materials Science}{32}{}{80-86}.
\newblock
\begin{APACrefDOI} \doi{10.1007/BF02538928} \end{APACrefDOI}
\PrintBackRefs{\CurrentBib}

\bibitem [\protect \citeauthoryear {%
Broberg%
}{%
Broberg%
}{%
{\protect \APACyear {1989}}%
{\protect \APACexlab {{\protect \BCnt {2}}}}}]{%
Broberg1989}
\APACinsertmetastar {%
Broberg1989}%
\begin{APACrefauthors}%
Broberg, K\BPBI B.%
\end{APACrefauthors}%
\unskip\
\newblock
\APACrefYearMonthDay{1989{\protect \BCnt {2}}}{}{}.
\newblock
{\BBOQ}\APACrefatitle {The near-tip field at high crack velocities} {The
  near-tip field at high crack velocities}.{\BBCQ}
\newblock
\APACjournalVolNumPages{International Journal of Fracture}{39}{}{}.
\newblock
\begin{APACrefDOI} \doi{10.1007/BF00047435} \end{APACrefDOI}
\PrintBackRefs{\CurrentBib}

\bibitem [\protect \citeauthoryear {%
Cicoira%
, Blatny%
, Li%
, Trottet%
\BCBL {}\ \BBA {} Gaume%
}{%
Cicoira%
\ \protect \BOthers {.}}{%
{\protect \APACyear {2022}}%
}]{%
cicoira2022}
\APACinsertmetastar {%
cicoira2022}%
\begin{APACrefauthors}%
Cicoira, A.%
, Blatny, L.%
, Li, X.%
, Trottet, B.%
\BCBL {}\ \BBA {} Gaume, J.%
\end{APACrefauthors}%
\unskip\
\newblock
\APACrefYearMonthDay{2022}{}{}.
\newblock
{\BBOQ}\APACrefatitle {Towards a predictive multi-phase model for alpine mass
  movements and process cascades} {Towards a predictive multi-phase model for
  alpine mass movements and process cascades}.{\BBCQ}
\newblock
\APACjournalVolNumPages{Engineering Geology}{310}{}{106866}.
\newblock
\begin{APACrefDOI} \doi{https://doi.org/10.1016/j.enggeo.2022.106866}
  \end{APACrefDOI}
\PrintBackRefs{\CurrentBib}

\bibitem [\protect \citeauthoryear {%
Daviet%
\ \BBA {} Bertails-Descoubes%
}{%
Daviet%
\ \BBA {} Bertails-Descoubes%
}{%
{\protect \APACyear {2016}}%
}]{%
Daviet2016}
\APACinsertmetastar {%
Daviet2016}%
\begin{APACrefauthors}%
Daviet, G.%
\BCBT {}\ \BBA {} Bertails-Descoubes, F.%
\end{APACrefauthors}%
\unskip\
\newblock
\APACrefYearMonthDay{2016}{}{}.
\newblock
{\BBOQ}\APACrefatitle {A Semi-Implicit Material Point Method for the Continuum
  Simulation of Granular Materials} {A semi-implicit material point method for
  the continuum simulation of granular materials}.{\BBCQ}
\newblock
\APACjournalVolNumPages{ACM Transactions on Graphics}{35}{}{}.
\newblock
\begin{APACrefDOI} \doi{10.1145/2897824.2925877} \end{APACrefDOI}
\PrintBackRefs{\CurrentBib}

\bibitem [\protect \citeauthoryear {%
de Souza~Neto%
, Peric%
\BCBL {}\ \BBA {} Owen%
}{%
de Souza~Neto%
\ \protect \BOthers {.}}{%
{\protect \APACyear {2008}}%
}]{%
souza}
\APACinsertmetastar {%
souza}%
\begin{APACrefauthors}%
de Souza~Neto, E\BPBI A.%
, Peric, D.%
\BCBL {}\ \BBA {} Owen, D\BPBI R\BPBI J.%
\end{APACrefauthors}%
\unskip\
\newblock
\APACrefYear{2008}.
\newblock
\APACrefbtitle {Computational {M}ethods for {P}lasticity} {Computational
  {M}ethods for {P}lasticity}.
\newblock
\APACaddressPublisher{}{Springer}.
\PrintBackRefs{\CurrentBib}

\bibitem [\protect \citeauthoryear {%
Dunatunga%
\ \BBA {} Kamrin%
}{%
Dunatunga%
\ \BBA {} Kamrin%
}{%
{\protect \APACyear {2015}}%
}]{%
Dunatunga_2015}
\APACinsertmetastar {%
Dunatunga_2015}%
\begin{APACrefauthors}%
Dunatunga, S.%
\BCBT {}\ \BBA {} Kamrin, K.%
\end{APACrefauthors}%
\unskip\
\newblock
\APACrefYearMonthDay{2015}{}{}.
\newblock
{\BBOQ}\APACrefatitle {Continuum modelling and simulation of granular flows
  through their many phases} {Continuum modelling and simulation of granular
  flows through their many phases}.{\BBCQ}
\newblock
\APACjournalVolNumPages{Journal of Fluid Mechanics}{779}{}{483--513}.
\newblock
\begin{APACrefDOI} \doi{10.1017/jfm.2015.383} \end{APACrefDOI}
\PrintBackRefs{\CurrentBib}

\bibitem [\protect \citeauthoryear {%
Failletaz%
, Louchet%
\BCBL {}\ \BBA {} Grasso%
}{%
Failletaz%
\ \protect \BOthers {.}}{%
{\protect \APACyear {2006}}%
}]{%
Failletaz2006}
\APACinsertmetastar {%
Failletaz2006}%
\begin{APACrefauthors}%
Failletaz, J.%
, Louchet, F.%
\BCBL {}\ \BBA {} Grasso, J\BHBI R.%
\end{APACrefauthors}%
\unskip\
\newblock
\APACrefYearMonthDay{2006}{}{}.
\newblock
{\BBOQ}\APACrefatitle {Cellular Automaton Modelling of Slab Avalanche
  Triggering Mechanisms: From the Universal Statistical Behaviour to Particular
  Cases} {Cellular automaton modelling of slab avalanche triggering mechanisms:
  From the universal statistical behaviour to particular cases}.{\BBCQ}
\newblock
\BIn{} \APACrefbtitle {International Snow Science Workshop} {International snow
  science workshop}\ (\BPG~174-180).
\newblock
\APACaddressPublisher{Telluride, Colorado}{}.
\PrintBackRefs{\CurrentBib}

\bibitem [\protect \citeauthoryear {%
Fyffe%
\ \BBA {} Zaiser%
}{%
Fyffe%
\ \BBA {} Zaiser%
}{%
{\protect \APACyear {2004}}%
}]{%
Fyffe2004}
\APACinsertmetastar {%
Fyffe2004}%
\begin{APACrefauthors}%
Fyffe, B.%
\BCBT {}\ \BBA {} Zaiser, M.%
\end{APACrefauthors}%
\unskip\
\newblock
\APACrefYearMonthDay{2004}{}{}.
\newblock
{\BBOQ}\APACrefatitle {The effects of snow variability on slab avalanche
  release} {The effects of snow variability on slab avalanche release}.{\BBCQ}
\newblock
\APACjournalVolNumPages{Cold Regions Science and Technology}{40}{}{229-242}.
\newblock
\begin{APACrefDOI} \doi{10.1016/j.coldregions.2004.08.004} \end{APACrefDOI}
\PrintBackRefs{\CurrentBib}

\bibitem [\protect \citeauthoryear {%
Fyffe%
\ \BBA {} Zaiser%
}{%
Fyffe%
\ \BBA {} Zaiser%
}{%
{\protect \APACyear {2007}}%
}]{%
Fyffe2007}
\APACinsertmetastar {%
Fyffe2007}%
\begin{APACrefauthors}%
Fyffe, B.%
\BCBT {}\ \BBA {} Zaiser, M.%
\end{APACrefauthors}%
\unskip\
\newblock
\APACrefYearMonthDay{2007}{}{}.
\newblock
{\BBOQ}\APACrefatitle {Interplay of basal shear fracture and slab rupture in
  slab avalanche release} {Interplay of basal shear fracture and slab rupture
  in slab avalanche release}.{\BBCQ}
\newblock
\APACjournalVolNumPages{Cold Regions Science and Technology}{49}{1}{26-38}.
\newblock
\APACrefnote{Selected Papers from the General Assembly of the European
  Geosciences Union (EGU), Vienna, Austria, 25 April 2005}
\newblock
\begin{APACrefDOI} \doi{10.1016/j.coldregions.2006.09.011} \end{APACrefDOI}
\PrintBackRefs{\CurrentBib}

\bibitem [\protect \citeauthoryear {%
Gao%
, Tampubolon%
, Jiang%
\BCBL {}\ \BBA {} Sifakis%
}{%
Gao%
\ \protect \BOthers {.}}{%
{\protect \APACyear {2017}}%
}]{%
Gao2017}
\APACinsertmetastar {%
Gao2017}%
\begin{APACrefauthors}%
Gao, M.%
, Tampubolon, A\BPBI P.%
, Jiang, C.%
\BCBL {}\ \BBA {} Sifakis, E.%
\end{APACrefauthors}%
\unskip\
\newblock
\APACrefYearMonthDay{2017}{}{}.
\newblock
{\BBOQ}\APACrefatitle {An Adaptive Generalized Interpolation Material Point
  Method for Simulating Elastoplastic Materials} {An adaptive generalized
  interpolation material point method for simulating elastoplastic
  materials}.{\BBCQ}
\newblock
\APACjournalVolNumPages{ACM Transactions on Graphics}{36}{6}{}.
\newblock
\begin{APACrefDOI} \doi{10.1145/3130800.3130879} \end{APACrefDOI}
\PrintBackRefs{\CurrentBib}

\bibitem [\protect \citeauthoryear {%
Gaume%
, Chambon%
, Eckert%
\BCBL {}\ \BBA {} Naaim%
}{%
Gaume%
\ \protect \BOthers {.}}{%
{\protect \APACyear {2013}}%
}]{%
gaume2013}
\APACinsertmetastar {%
gaume2013}%
\begin{APACrefauthors}%
Gaume, J.%
, Chambon, G.%
, Eckert, N.%
\BCBL {}\ \BBA {} Naaim, M.%
\end{APACrefauthors}%
\unskip\
\newblock
\APACrefYearMonthDay{2013}{}{}.
\newblock
{\BBOQ}\APACrefatitle {Influence of weak-layer heterogeneity on snow slab
  avalanche release: application to the evaluation of avalanche release depths}
  {Influence of weak-layer heterogeneity on snow slab avalanche release:
  application to the evaluation of avalanche release depths}.{\BBCQ}
\newblock
\APACjournalVolNumPages{Journal of Glaciology}{59}{215}{423–437}.
\newblock
\begin{APACrefDOI} \doi{10.3189/2013JoG12J161} \end{APACrefDOI}
\PrintBackRefs{\CurrentBib}

\bibitem [\protect \citeauthoryear {%
Gaume%
, Chambon%
, Eckert%
, Naaim%
\BCBL {}\ \BBA {} Schweizer%
}{%
Gaume%
, Chambon%
\BCBL {}\ \protect \BOthers {.}}{%
{\protect \APACyear {2015}}%
}]{%
Gaume2015a}
\APACinsertmetastar {%
Gaume2015a}%
\begin{APACrefauthors}%
Gaume, J.%
, Chambon, G.%
, Eckert, N.%
, Naaim, M.%
\BCBL {}\ \BBA {} Schweizer, J.%
\end{APACrefauthors}%
\unskip\
\newblock
\APACrefYearMonthDay{2015}{}{}.
\newblock
{\BBOQ}\APACrefatitle {Influence of weak layer heterogeneity and slab
  properties on slab tensile failure propensity and avalanche release area}
  {Influence of weak layer heterogeneity and slab properties on slab tensile
  failure propensity and avalanche release area}.{\BBCQ}
\newblock
\APACjournalVolNumPages{The Cryosphere}{9}{2}{795--804}.
\newblock
\begin{APACrefDOI} \doi{10.5194/tc-9-795-2015} \end{APACrefDOI}
\PrintBackRefs{\CurrentBib}

\bibitem [\protect \citeauthoryear {%
Gaume%
, Gast%
, Terran%
, Van~Herwinjen%
\BCBL {}\ \BBA {} Jiang%
}{%
Gaume%
\ \protect \BOthers {.}}{%
{\protect \APACyear {2018}}%
}]{%
Gaume2018}
\APACinsertmetastar {%
Gaume2018}%
\begin{APACrefauthors}%
Gaume, J.%
, Gast, T.%
, Terran, J.%
, Van~Herwinjen, A.%
\BCBL {}\ \BBA {} Jiang, C.%
\end{APACrefauthors}%
\unskip\
\newblock
\APACrefYearMonthDay{2018}{}{}.
\newblock
{\BBOQ}\APACrefatitle {Dynamic anticrack propagation in snow} {Dynamic
  anticrack propagation in snow}.{\BBCQ}
\newblock
\APACjournalVolNumPages{Nature Communication}{9}{3047}{}.
\newblock
\begin{APACrefDOI} \doi{10.1038/s41467-018-05181-w} \end{APACrefDOI}
\PrintBackRefs{\CurrentBib}

\bibitem [\protect \citeauthoryear {%
Gaume%
\ \BBA {} Reuter%
}{%
Gaume%
\ \BBA {} Reuter%
}{%
{\protect \APACyear {2017}}%
}]{%
GaumeReuter2017}
\APACinsertmetastar {%
GaumeReuter2017}%
\begin{APACrefauthors}%
Gaume, J.%
\BCBT {}\ \BBA {} Reuter, B.%
\end{APACrefauthors}%
\unskip\
\newblock
\APACrefYearMonthDay{2017}{}{}.
\newblock
{\BBOQ}\APACrefatitle {Assessing snow instability in skier-triggered snow slab
  avalanches by combining failure initiation and crack propagation} {Assessing
  snow instability in skier-triggered snow slab avalanches by combining failure
  initiation and crack propagation}.{\BBCQ}
\newblock
\APACjournalVolNumPages{Cold regions science and technology}{144}{}{6--15}.
\newblock
\APACrefnote{International Snow Science Workshop 2016 Breckenridge}
\newblock
\begin{APACrefDOI} \doi{https://doi.org/10.1016/j.coldregions.2017.05.011}
  \end{APACrefDOI}
\PrintBackRefs{\CurrentBib}

\bibitem [\protect \citeauthoryear {%
Gaume%
, {van Herwijnen}%
, Gast%
, Teran%
\BCBL {}\ \BBA {} Jiang%
}{%
Gaume%
\ \protect \BOthers {.}}{%
{\protect \APACyear {2019}}%
}]{%
Gaume2019}
\APACinsertmetastar {%
Gaume2019}%
\begin{APACrefauthors}%
Gaume, J.%
, {van Herwijnen}, A.%
, Gast, T.%
, Teran, J.%
\BCBL {}\ \BBA {} Jiang, C.%
\end{APACrefauthors}%
\unskip\
\newblock
\APACrefYearMonthDay{2019}{}{}.
\newblock
{\BBOQ}\APACrefatitle {Investigating the release and flow of snow avalanches at
  the slope-scale using a unified model based on the material point method}
  {Investigating the release and flow of snow avalanches at the slope-scale
  using a unified model based on the material point method}.{\BBCQ}
\newblock
\APACjournalVolNumPages{Cold Regions Science and Technology}{168}{}{102847}.
\newblock
\begin{APACrefDOI} \doi{10.1016/j.coldregions.2019.102847} \end{APACrefDOI}
\PrintBackRefs{\CurrentBib}

\bibitem [\protect \citeauthoryear {%
Gaume%
, van Herwijnen%
, Chambon%
, Birkeland%
\BCBL {}\ \BBA {} Schweizer%
}{%
Gaume%
, van Herwijnen%
\BCBL {}\ \protect \BOthers {.}}{%
{\protect \APACyear {2015}}%
}]{%
Gaume2015}
\APACinsertmetastar {%
Gaume2015}%
\begin{APACrefauthors}%
Gaume, J.%
, van Herwijnen, A.%
, Chambon, G.%
, Birkeland, K\BPBI W.%
\BCBL {}\ \BBA {} Schweizer, J.%
\end{APACrefauthors}%
\unskip\
\newblock
\APACrefYearMonthDay{2015}{}{}.
\newblock
{\BBOQ}\APACrefatitle {Modeling of crack propagation in weak snowpack layers
  using the discrete element method} {Modeling of crack propagation in weak
  snowpack layers using the discrete element method}.{\BBCQ}
\newblock
\APACjournalVolNumPages{The Cryosphere}{9}{5}{1915--1932}.
\newblock
\begin{APACrefDOI} \doi{10.5194/tc-9-1915-2015} \end{APACrefDOI}
\PrintBackRefs{\CurrentBib}

\bibitem [\protect \citeauthoryear {%
Jiang%
, Schroeder%
, Teran%
, Stomakhin%
\BCBL {}\ \BBA {} Selle%
}{%
Jiang%
\ \protect \BOthers {.}}{%
{\protect \APACyear {2016}}%
}]{%
JIANG}
\APACinsertmetastar {%
JIANG}%
\begin{APACrefauthors}%
Jiang, C.%
, Schroeder, C.%
, Teran, J.%
, Stomakhin, A.%
\BCBL {}\ \BBA {} Selle, A.%
\end{APACrefauthors}%
\unskip\
\newblock
\APACrefYearMonthDay{2016}{}{}.
\newblock
{\BBOQ}\APACrefatitle {The Material Point Method for Simulating Continuum
  Materials} {The material point method for simulating continuum
  materials}.{\BBCQ}
\newblock
\BIn{} \APACrefbtitle {ACM SIGGRAPH 2016 Courses.} {Acm siggraph 2016 courses.}
\newblock
\APACaddressPublisher{New York, NY, USA}{Association for Computing Machinery}.
\PrintBackRefs{\CurrentBib}

\bibitem [\protect \citeauthoryear {%
Li%
, Sovilla%
, Jiang%
\BCBL {}\ \BBA {} Gaume%
}{%
Li%
\ \protect \BOthers {.}}{%
{\protect \APACyear {2021}}%
}]{%
Li2021}
\APACinsertmetastar {%
Li2021}%
\begin{APACrefauthors}%
Li, X.%
, Sovilla, B.%
, Jiang, C.%
\BCBL {}\ \BBA {} Gaume, J.%
\end{APACrefauthors}%
\unskip\
\newblock
\APACrefYearMonthDay{2021}{}{}.
\newblock
{\BBOQ}\APACrefatitle {Three-dimensional and real-scale modeling of flow
  regimes in dense snow avalanches} {Three-dimensional and real-scale modeling
  of flow regimes in dense snow avalanches}.{\BBCQ}
\newblock
\APACjournalVolNumPages{Landslides}{18}{10}{3393--3406}.
\newblock
\begin{APACrefDOI} \doi{https://doi.org/10.1007/s10346-021-01692-8}
  \end{APACrefDOI}
\PrintBackRefs{\CurrentBib}

\bibitem [\protect \citeauthoryear {%
Mahajan%
, Kalakuntla%
\BCBL {}\ \BBA {} Chandel%
}{%
Mahajan%
\ \protect \BOthers {.}}{%
{\protect \APACyear {2010}}%
}]{%
Mahajan2010}
\APACinsertmetastar {%
Mahajan2010}%
\begin{APACrefauthors}%
Mahajan, P.%
, Kalakuntla, R.%
\BCBL {}\ \BBA {} Chandel, C.%
\end{APACrefauthors}%
\unskip\
\newblock
\APACrefYearMonthDay{2010}{}{}.
\newblock
{\BBOQ}\APACrefatitle {Numerical simulation of failure in a layered thin
  snowpack under skier load} {Numerical simulation of failure in a layered thin
  snowpack under skier load}.{\BBCQ}
\newblock
\APACjournalVolNumPages{Annals of Glaciology}{51}{54}{169–175}.
\newblock
\begin{APACrefDOI} \doi{10.3189/172756410791386436} \end{APACrefDOI}
\PrintBackRefs{\CurrentBib}

\bibitem [\protect \citeauthoryear {%
D.~McClung%
\ \BBA {} Schaerer%
}{%
D.~McClung%
\ \BBA {} Schaerer%
}{%
{\protect \APACyear {2006}}%
}]{%
Mcclung2006}
\APACinsertmetastar {%
Mcclung2006}%
\begin{APACrefauthors}%
McClung, D.%
\BCBT {}\ \BBA {} Schaerer, P\BPBI A.%
\end{APACrefauthors}%
\unskip\
\newblock
\APACrefYear{2006}.
\newblock
\APACrefbtitle {The avalanche handbook} {The avalanche handbook}.
\newblock
\APACaddressPublisher{}{The Mountaineers Books}.
\PrintBackRefs{\CurrentBib}

\bibitem [\protect \citeauthoryear {%
D\BPBI M.~McClung%
}{%
D\BPBI M.~McClung%
}{%
{\protect \APACyear {2003}}%
}]{%
Mclung2003}
\APACinsertmetastar {%
Mclung2003}%
\begin{APACrefauthors}%
McClung, D\BPBI M.%
\end{APACrefauthors}%
\unskip\
\newblock
\APACrefYearMonthDay{2003}{}{}.
\newblock
{\BBOQ}\APACrefatitle {Size scaling for dry snow slab release} {Size scaling
  for dry snow slab release}.{\BBCQ}
\newblock
\APACjournalVolNumPages{Journal of Geophysical Research: Solid
  Earth}{108}{B10}{}.
\newblock
\begin{APACrefDOI} \doi{10.1029/2002JB002298} \end{APACrefDOI}
\PrintBackRefs{\CurrentBib}

\bibitem [\protect \citeauthoryear {%
D\BPBI M.~McClung%
}{%
D\BPBI M.~McClung%
}{%
{\protect \APACyear {2009}}%
}]{%
McClung2009}
\APACinsertmetastar {%
McClung2009}%
\begin{APACrefauthors}%
McClung, D\BPBI M.%
\end{APACrefauthors}%
\unskip\
\newblock
\APACrefYearMonthDay{2009}{}{}.
\newblock
{\BBOQ}\APACrefatitle {Dimensions of dry snow slab avalanches from field
  measurements} {Dimensions of dry snow slab avalanches from field
  measurements}.{\BBCQ}
\newblock
\APACjournalVolNumPages{Journal of Geophysical Research: Earth
  Surface}{114}{F1}{}.
\newblock
\begin{APACrefDOI} \doi{10.1029/2007JF000941} \end{APACrefDOI}
\PrintBackRefs{\CurrentBib}

\bibitem [\protect \citeauthoryear {%
Meschke%
, Liu%
\BCBL {}\ \BBA {} Mang%
}{%
Meschke%
\ \protect \BOthers {.}}{%
{\protect \APACyear {1996}}%
}]{%
Meschke1996}
\APACinsertmetastar {%
Meschke1996}%
\begin{APACrefauthors}%
Meschke, G.%
, Liu, C.%
\BCBL {}\ \BBA {} Mang, H\BPBI A.%
\end{APACrefauthors}%
\unskip\
\newblock
\APACrefYearMonthDay{1996}{}{}.
\newblock
{\BBOQ}\APACrefatitle {Large Strain Finite-Element Analysis of Snow} {Large
  strain finite-element analysis of snow}.{\BBCQ}
\newblock
\APACjournalVolNumPages{Journal of Engineering Mechanics}{122}{7}{591-602}.
\newblock
\begin{APACrefDOI} \doi{10.1061/(ASCE)0733-9399(1996)122:7(591)}
  \end{APACrefDOI}
\PrintBackRefs{\CurrentBib}

\bibitem [\protect \citeauthoryear {%
Reuter%
\ \BBA {} Schweizer%
}{%
Reuter%
\ \BBA {} Schweizer%
}{%
{\protect \APACyear {2018}}%
}]{%
Reuter2018}
\APACinsertmetastar {%
Reuter2018}%
\begin{APACrefauthors}%
Reuter, B.%
\BCBT {}\ \BBA {} Schweizer, J.%
\end{APACrefauthors}%
\unskip\
\newblock
\APACrefYearMonthDay{2018}{}{}.
\newblock
{\BBOQ}\APACrefatitle {Describing snow instability by failure initiation, crack
  propagation, and slab tensile support} {Describing snow instability by
  failure initiation, crack propagation, and slab tensile support}.{\BBCQ}
\newblock
\APACjournalVolNumPages{Geophysical Research Letters}{45}{14}{7019--7027}.
\newblock
\begin{APACrefDOI} \doi{https://doi.org/10.1029/2018GL078069} \end{APACrefDOI}
\PrintBackRefs{\CurrentBib}

\bibitem [\protect \citeauthoryear {%
Roscoe%
\ \BBA {} Burland%
}{%
Roscoe%
\ \BBA {} Burland%
}{%
{\protect \APACyear {1968}}%
}]{%
Roscoe1968}
\APACinsertmetastar {%
Roscoe1968}%
\begin{APACrefauthors}%
Roscoe, K.%
\BCBT {}\ \BBA {} Burland, J.%
\end{APACrefauthors}%
\unskip\
\newblock
\APACrefYearMonthDay{1968}{}{}.
\newblock
{\BBOQ}\APACrefatitle {On the Generalized Stress-Strain Behavior of Wet Clays}
  {On the generalized stress-strain behavior of wet clays}.{\BBCQ}
\newblock
\BIn{} J.~Heyman\ \BBA {} F.~Leckie\ (\BEDS), \APACrefbtitle {Engineering
  Plasticity.} {Engineering plasticity.}
\newblock
\APACaddressPublisher{}{Cambridge University Press}.
\PrintBackRefs{\CurrentBib}

\bibitem [\protect \citeauthoryear {%
Savage%
\ \BBA {} Huter%
}{%
Savage%
\ \BBA {} Huter%
}{%
{\protect \APACyear {1915}}%
}]{%
Savage_Hutter}
\APACinsertmetastar {%
Savage_Hutter}%
\begin{APACrefauthors}%
Savage, S\BPBI B.%
\BCBT {}\ \BBA {} Huter, K.%
\end{APACrefauthors}%
\unskip\
\newblock
\APACrefYearMonthDay{1915}{}{}.
\newblock
{\BBOQ}\APACrefatitle {The dynamics of avalanches of granular materials from
  initiation to runout. Part I: Analysis} {The dynamics of avalanches of
  granular materials from initiation to runout. part i: Analysis}.{\BBCQ}
\newblock
\APACjournalVolNumPages{Acta Mechanica}{86}{}{201-223}.
\newblock
\APACrefnote{Particle Simulation Methods}
\newblock
\begin{APACrefDOI} \doi{10.1007/BF01175958} \end{APACrefDOI}
\PrintBackRefs{\CurrentBib}

\bibitem [\protect \citeauthoryear {%
Schreck%
\ \BBA {} Wojtan%
}{%
Schreck%
\ \BBA {} Wojtan%
}{%
{\protect \APACyear {2020}}%
}]{%
SCHRECK2020}
\APACinsertmetastar {%
SCHRECK2020}%
\begin{APACrefauthors}%
Schreck, C.%
\BCBT {}\ \BBA {} Wojtan, C.%
\end{APACrefauthors}%
\unskip\
\newblock
\APACrefYearMonthDay{2020}{}{}.
\newblock
{\BBOQ}\APACrefatitle {A Practical Method for Animating Anisotropic
  Elastoplastic Materials} {A practical method for animating anisotropic
  elastoplastic materials}.{\BBCQ}
\newblock
\APACjournalVolNumPages{Computer Graphics Forum}{39}{2}{89-99}.
\newblock
\begin{APACrefDOI} \doi{10.1111/cgf.13914} \end{APACrefDOI}
\PrintBackRefs{\CurrentBib}

\bibitem [\protect \citeauthoryear {%
Schweizer%
, Bruce~Jamieson%
\BCBL {}\ \BBA {} Schneebeli%
}{%
Schweizer%
\ \protect \BOthers {.}}{%
{\protect \APACyear {2003}}%
}]{%
Schweizer2003}
\APACinsertmetastar {%
Schweizer2003}%
\begin{APACrefauthors}%
Schweizer, J.%
, Bruce~Jamieson, J.%
\BCBL {}\ \BBA {} Schneebeli, M.%
\end{APACrefauthors}%
\unskip\
\newblock
\APACrefYearMonthDay{2003}{}{}.
\newblock
{\BBOQ}\APACrefatitle {Snow avalanche formation} {Snow avalanche
  formation}.{\BBCQ}
\newblock
\APACjournalVolNumPages{Reviews of Geophysics}{41}{4}{}.
\newblock
\begin{APACrefDOI} \doi{10.1029/2002RG000123} \end{APACrefDOI}
\PrintBackRefs{\CurrentBib}

\bibitem [\protect \citeauthoryear {%
Schweizer%
, Reuter%
, van Herwijnen%
\BCBL {}\ \BBA {} Gaume%
}{%
Schweizer%
\ \protect \BOthers {.}}{%
{\protect \APACyear {2016}}%
}]{%
Schweizer2016}
\APACinsertmetastar {%
Schweizer2016}%
\begin{APACrefauthors}%
Schweizer, J.%
, Reuter, B.%
, van Herwijnen, A.%
\BCBL {}\ \BBA {} Gaume, J.%
\end{APACrefauthors}%
\unskip\
\newblock
\APACrefYearMonthDay{2016}{}{}.
\newblock
{\BBOQ}\APACrefatitle {Avalanche Release 101} {Avalanche release 101}.{\BBCQ}
\newblock
\BIn{} \APACrefbtitle {International Snow Science Workshop} {International snow
  science workshop}\ (\BPG~1-11).
\newblock
\APACaddressPublisher{Breckenridge, Colorado}{}.
\PrintBackRefs{\CurrentBib}

\bibitem [\protect \citeauthoryear {%
Soga%
, Alonso%
, Yerro%
, Kumar%
\BCBL {}\ \BBA {} Bandara%
}{%
Soga%
\ \protect \BOthers {.}}{%
{\protect \APACyear {2016}}%
}]{%
soga2016}
\APACinsertmetastar {%
soga2016}%
\begin{APACrefauthors}%
Soga, K.%
, Alonso, E.%
, Yerro, A.%
, Kumar, K.%
\BCBL {}\ \BBA {} Bandara, S.%
\end{APACrefauthors}%
\unskip\
\newblock
\APACrefYearMonthDay{2016}{}{}.
\newblock
{\BBOQ}\APACrefatitle {Trends in large-deformation analysis of landslide mass
  movements with particular emphasis on the material point method} {Trends in
  large-deformation analysis of landslide mass movements with particular
  emphasis on the material point method}.{\BBCQ}
\newblock
\APACjournalVolNumPages{Géotechnique}{66}{3}{248-273}.
\newblock
\begin{APACrefDOI} \doi{10.1680/jgeot.15.LM.005} \end{APACrefDOI}
\PrintBackRefs{\CurrentBib}

\bibitem [\protect \citeauthoryear {%
Stomakhin%
, Schroeder%
, Chai%
, Teran%
\BCBL {}\ \BBA {} Selle%
}{%
Stomakhin%
\ \protect \BOthers {.}}{%
{\protect \APACyear {2013}}%
}]{%
stomakhin}
\APACinsertmetastar {%
stomakhin}%
\begin{APACrefauthors}%
Stomakhin, A.%
, Schroeder, C.%
, Chai, L.%
, Teran, J.%
\BCBL {}\ \BBA {} Selle, A.%
\end{APACrefauthors}%
\unskip\
\newblock
\APACrefYearMonthDay{2013}{}{}.
\newblock
{\BBOQ}\APACrefatitle {A Material Point Method for Snow Simulation} {A material
  point method for snow simulation}.{\BBCQ}
\newblock
\APACjournalVolNumPages{ACM Trans. Graph.}{32}{4}{}.
\newblock
\begin{APACrefDOI} \doi{10.1145/2461912.2461948} \end{APACrefDOI}
\PrintBackRefs{\CurrentBib}

\bibitem [\protect \citeauthoryear {%
Sulsky%
, Chen%
\BCBL {}\ \BBA {} Schreyer%
}{%
Sulsky%
\ \protect \BOthers {.}}{%
{\protect \APACyear {1994}}%
}]{%
SULSKY1994}
\APACinsertmetastar {%
SULSKY1994}%
\begin{APACrefauthors}%
Sulsky, D.%
, Chen, Z.%
\BCBL {}\ \BBA {} Schreyer, H.%
\end{APACrefauthors}%
\unskip\
\newblock
\APACrefYearMonthDay{1994}{}{}.
\newblock
{\BBOQ}\APACrefatitle {A particle method for history-dependent materials} {A
  particle method for history-dependent materials}.{\BBCQ}
\newblock
\APACjournalVolNumPages{Computer Methods in Applied Mechanics and
  Engineering}{118}{1}{179-196}.
\newblock
\begin{APACrefDOI} \doi{10.1016/0045-7825(94)90112-0} \end{APACrefDOI}
\PrintBackRefs{\CurrentBib}

\bibitem [\protect \citeauthoryear {%
Sulsky%
\ \BBA {} Peterson%
}{%
Sulsky%
\ \BBA {} Peterson%
}{%
{\protect \APACyear {2011}}%
}]{%
SULSKY20111674}
\APACinsertmetastar {%
SULSKY20111674}%
\begin{APACrefauthors}%
Sulsky, D.%
\BCBT {}\ \BBA {} Peterson, K.%
\end{APACrefauthors}%
\unskip\
\newblock
\APACrefYearMonthDay{2011}{}{}.
\newblock
{\BBOQ}\APACrefatitle {Toward a new elastic–decohesive model of Arctic sea
  ice} {Toward a new elastic–decohesive model of arctic sea ice}.{\BBCQ}
\newblock
\APACjournalVolNumPages{Physica D: Nonlinear Phenomena}{240}{20}{1674-1683}.
\newblock
\APACrefnote{Special Issue: Fluid Dynamics: From Theory to Experiment}
\newblock
\begin{APACrefDOI} \doi{10.1016/j.physd.2011.07.005} \end{APACrefDOI}
\PrintBackRefs{\CurrentBib}

\bibitem [\protect \citeauthoryear {%
Sulsky%
, Zhou%
\BCBL {}\ \BBA {} Schreyer%
}{%
Sulsky%
\ \protect \BOthers {.}}{%
{\protect \APACyear {1995}}%
}]{%
SULSKY1995}
\APACinsertmetastar {%
SULSKY1995}%
\begin{APACrefauthors}%
Sulsky, D.%
, Zhou, S\BHBI J.%
\BCBL {}\ \BBA {} Schreyer, H.%
\end{APACrefauthors}%
\unskip\
\newblock
\APACrefYearMonthDay{1995}{}{}.
\newblock
{\BBOQ}\APACrefatitle {Application of a particle-in-cell method to solid
  mechanics} {Application of a particle-in-cell method to solid
  mechanics}.{\BBCQ}
\newblock
\APACjournalVolNumPages{Computer Physics Communications}{87}{1}{236-252}.
\newblock
\APACrefnote{Particle Simulation Methods}
\newblock
\begin{APACrefDOI} \doi{10.1016/0010-4655(94)00170-7} \end{APACrefDOI}
\PrintBackRefs{\CurrentBib}

\bibitem [\protect \citeauthoryear {%
Trottet%
\ \protect \BOthers {.}}{%
Trottet%
\ \protect \BOthers {.}}{%
{\protect \APACyear {2022}}%
}]{%
Trottet2022}
\APACinsertmetastar {%
Trottet2022}%
\begin{APACrefauthors}%
Trottet, B.%
, Simenhois, R.%
, Bobillier, G.%
, Bergfeld, B.%
, {van Herwijnen}, A.%
, Jiang, C.%
\BCBL {}\ \BBA {} Gaume, J.%
\end{APACrefauthors}%
\unskip\
\newblock
\APACrefYearMonthDay{2022}{}{}.
\newblock
{\BBOQ}\APACrefatitle {Transition from sub-Rayleigh anticrack to supershear
  crack propagation in snow avalanches} {Transition from sub-rayleigh anticrack
  to supershear crack propagation in snow avalanches}.{\BBCQ}
\newblock
\APACjournalVolNumPages{Nature Physics}{18}{}{1094-1098}.
\newblock
\begin{APACrefDOI} \doi{10.1038/s41567-022-01662-4} \end{APACrefDOI}
\PrintBackRefs{\CurrentBib}

\bibitem [\protect \citeauthoryear {%
van Herwijnen%
\ \protect \BOthers {.}}{%
van Herwijnen%
\ \protect \BOthers {.}}{%
{\protect \APACyear {2016}}%
}]{%
Alec2016}
\APACinsertmetastar {%
Alec2016}%
\begin{APACrefauthors}%
van Herwijnen, A.%
, Bair, E\BPBI H.%
, Birkeland, K\BPBI W.%
, Reuter, B.%
, Simenhois, R.%
, Jamieson, B.%
\BCBL {}\ \BBA {} Schweizer, J.%
\end{APACrefauthors}%
\unskip\
\newblock
\APACrefYearMonthDay{2016}{}{}.
\newblock
{\BBOQ}\APACrefatitle {Measuring the mechanical properties of snow relevant for
  dry-snow slab avalanche release using particle tracking velocimetry}
  {Measuring the mechanical properties of snow relevant for dry-snow slab
  avalanche release using particle tracking velocimetry}.{\BBCQ}
\newblock
\BIn{} \APACrefbtitle {International Snow Science Workshop} {International snow
  science workshop}\ (\BPG~397-404).
\newblock
\APACaddressPublisher{Breckenridge, Colorado}{}.
\PrintBackRefs{\CurrentBib}

\bibitem [\protect \citeauthoryear {%
Veitinger%
, Purves%
\BCBL {}\ \BBA {} Sovilla%
}{%
Veitinger%
\ \protect \BOthers {.}}{%
{\protect \APACyear {2016}}%
}]{%
veitinger2016}
\APACinsertmetastar {%
veitinger2016}%
\begin{APACrefauthors}%
Veitinger, J.%
, Purves, R\BPBI S.%
\BCBL {}\ \BBA {} Sovilla, B.%
\end{APACrefauthors}%
\unskip\
\newblock
\APACrefYearMonthDay{2016}{}{}.
\newblock
{\BBOQ}\APACrefatitle {Potential slab avalanche release area identification
  from estimated winter terrain: a multi-scale, fuzzy logic approach}
  {Potential slab avalanche release area identification from estimated winter
  terrain: a multi-scale, fuzzy logic approach}.{\BBCQ}
\newblock
\APACjournalVolNumPages{Natural Hazards and Earth System
  Sciences}{16}{10}{2211--2225}.
\newblock
\begin{APACrefDOI} \doi{10.5194/nhess-16-2211-2016} \end{APACrefDOI}
\PrintBackRefs{\CurrentBib}

\bibitem [\protect \citeauthoryear {%
Vicari%
, Tran%
, Nordal%
\BCBL {}\ \BBA {} Thakur%
}{%
Vicari%
\ \protect \BOthers {.}}{%
{\protect \APACyear {2022}}%
}]{%
vicari2022}
\APACinsertmetastar {%
vicari2022}%
\begin{APACrefauthors}%
Vicari, H.%
, Tran, Q\BPBI A.%
, Nordal, S.%
\BCBL {}\ \BBA {} Thakur, V.%
\end{APACrefauthors}%
\unskip\
\newblock
\APACrefYearMonthDay{2022}{}{}.
\newblock
{\BBOQ}\APACrefatitle {MPM modelling of debris flow entrainment and interaction
  with an upstream flexible barrier} {Mpm modelling of debris flow entrainment
  and interaction with an upstream flexible barrier}.{\BBCQ}
\newblock
\APACjournalVolNumPages{Landslides}{}{}{1--15}.
\newblock
\begin{APACrefDOI} \doi{https://doi.org/10.1007/s10346-022-01886-8}
  \end{APACrefDOI}
\PrintBackRefs{\CurrentBib}

\bibitem [\protect \citeauthoryear {%
Wolper%
\ \protect \BOthers {.}}{%
Wolper%
\ \protect \BOthers {.}}{%
{\protect \APACyear {2021}}%
}]{%
wolper2021}
\APACinsertmetastar {%
wolper2021}%
\begin{APACrefauthors}%
Wolper, J.%
, Gao, M.%
, L{\"u}thi, M\BPBI P.%
, Heller, V.%
, Vieli, A.%
, Jiang, C.%
\BCBL {}\ \BBA {} Gaume, J.%
\end{APACrefauthors}%
\unskip\
\newblock
\APACrefYearMonthDay{2021}{}{}.
\newblock
{\BBOQ}\APACrefatitle {A glacier--ocean interaction model for tsunami genesis
  due to iceberg calving} {A glacier--ocean interaction model for tsunami
  genesis due to iceberg calving}.{\BBCQ}
\newblock
\APACjournalVolNumPages{Communications Earth \& Environment}{2}{1}{1--10}.
\newblock
\begin{APACrefDOI} \doi{https://doi.org/10.1038/s43247-021-00179-7}
  \end{APACrefDOI}
\PrintBackRefs{\CurrentBib}

\bibitem [\protect \citeauthoryear {%
York~II%
, Sulsky%
\BCBL {}\ \BBA {} Schreyer%
}{%
York~II%
\ \protect \BOthers {.}}{%
{\protect \APACyear {2000}}%
}]{%
york1999}
\APACinsertmetastar {%
york1999}%
\begin{APACrefauthors}%
York~II, A\BPBI R.%
, Sulsky, D.%
\BCBL {}\ \BBA {} Schreyer, H\BPBI L.%
\end{APACrefauthors}%
\unskip\
\newblock
\APACrefYearMonthDay{2000}{}{}.
\newblock
{\BBOQ}\APACrefatitle {Fluid–membrane interaction based on the material point
  method} {Fluid–membrane interaction based on the material point
  method}.{\BBCQ}
\newblock
\APACjournalVolNumPages{International Journal for Numerical Methods in
  Engineering}{48}{6}{901-924}.
\newblock
\begin{APACrefDOI}
  \doi{10.1002/(SICI)1097-0207(20000630)48:6<901::AID-NME910>3.0.CO;2-T}
  \end{APACrefDOI}
\PrintBackRefs{\CurrentBib}

\bibitem [\protect \citeauthoryear {%
Zhang%
\ \BBA {} Puzrin%
}{%
Zhang%
\ \BBA {} Puzrin%
}{%
{\protect \APACyear {2022}}%
}]{%
Zhang2022}
\APACinsertmetastar {%
Zhang2022}%
\begin{APACrefauthors}%
Zhang, W.%
\BCBT {}\ \BBA {} Puzrin, A\BPBI M.%
\end{APACrefauthors}%
\unskip\
\newblock
\APACrefYearMonthDay{2022}{}{}.
\newblock
{\BBOQ}\APACrefatitle {How Small Slip Surfaces Evolve Into Large Submarine
  Landslides—Insight From 3D Numerical Modeling} {How small slip surfaces
  evolve into large submarine landslides—insight from 3d numerical
  modeling}.{\BBCQ}
\newblock
\APACjournalVolNumPages{Journal of Geophysical Research: Earth
  Surface}{127}{7}{e2022JF006640}.
\newblock
\begin{APACrefDOI} \doi{10.1029/2022JF006640} \end{APACrefDOI}
\PrintBackRefs{\CurrentBib}

\end{thebibliography}
\end{document}